\documentclass[aip,cha,reprint]{revtex4-1}

\usepackage{graphicx}
\usepackage{dcolumn}
\usepackage{amsmath}
\usepackage{amssymb}
\usepackage{bm}
\usepackage{amsbsy}
\usepackage[utf8]{inputenc}
\usepackage[T1]{fontenc}
\usepackage{mathptmx}
\usepackage{etoolbox}
\usepackage{mathtools}
\usepackage{color,soul}
\usepackage{float}
\graphicspath{{Figures/}}
\usepackage{subfigure}
\DeclarePairedDelimiter\abs{\lvert}{\rvert}
\DeclarePairedDelimiter\norm{\lVert}{\rVert}
\DeclarePairedDelimiter{\innerproduct}{\langle}{\rangle}

\makeatletter
\def\@email#1#2{%
 \endgroup
 \patchcmd{\titleblock@produce}
  {\frontmatter@RRAPformat}
  {\frontmatter@RRAPformat{\produce@RRAP{*#1\href{mailto:#2}{#2}}}\frontmatter@RRAPformat}
  {}{}
}%
\makeatother

\begin{document}

\title[]{A unified framework for synchronization optimization in directed multiplex networks}

\author{Anath Bandhu Das}
\email{anathbandhu0498@gmail.com}

\author{Pinaki Pal}%
\email{ppal.maths@nitdgp.ac.in (Corresponding Author)}

\affiliation{Department of Mathematics, National Institute of Technology, Durgapur 713209, India}%

\date{\today}

\begin{abstract}
The multiplex network paradigm has been instrumental in revealing many unexpected phenomena and dynamical regimes in complex interacting systems. Nevertheless, most of the current research focuses on undirected multiplex structures, whereas real-world systems predominantly involve directed interactions. Here, we present an analytical framework for attaining optimal synchronization in directed multiplex networks composed of phase oscillators, considering both frustrated and non-frustrated regimes. A multiplex synchrony alignment function (MSAF) is introduced for this purpose, whose formulation integrates structural properties and dynamical characteristics of the individual directed layers. Using this function, we derive two classes of frequency distributions: one that yields perfect synchronization at a prescribed coupling strength in the presence of phase-lag, and another that optimizes synchronization over a broad range of coupling strengths. Numerical simulations on various directed duplex topologies demonstrate that both frequency sets substantially outperform conventional distributions. We also explore network optimization through a directed link rewiring strategy aimed at minimizing the MSAF, along with a swapping algorithm for optimally assigning fixed frequencies on both layers of a given directed duplex network. Examination of synchrony-optimized directed networks uncovers three notable correlations: a positive relationship between frequency and out-degree, a negative correlation between neighboring frequencies, and an anti-correlation between mirror node frequencies across directed layers.
\end{abstract}

\maketitle

\begin{quotation}
The study of complex networks has provided profound insights into diverse phenomena, from disease spreading and biological pathways to engineered control systems. Yet, real-world networks rarely exist in isolation; they are interconnected, forming multilayer systems where interactions often exhibit directionality. A prime example of a directed multiplex network is the global financial system, where transactions flow between institutions across different layers such as interbank lending, derivatives trading, and foreign exchange markets---each layer having its own directed connectivity patterns that influence systemic stability. Similarly, in neural systems, distinct brain regions communicate via directed synaptic pathways across multiple frequency bands, forming a directed multiplex architecture critical for cognitive function. Observations from single-network analyses often fail to capture the emergent dynamics that arise when networks are coupled, underscoring the necessity of a directed multiplex framework. In recent years, synchronization has emerged as a central topic of investigation within multiplex networks, driven by its relevance to both fundamental science and practical applications. In this work, we develop an analytical approach for achieving optimal synchronization in directed duplex networks of phase oscillators. By introducing a multiplex synchronization alignment function that encodes both structural and dynamical information from each layer, we derive two frequency sets: one enabling perfect synchronization at a targeted coupling strength and another enhancing synchronization over a broad range. We further address network design through simultaneous rewiring of both layers and optimal frequency arrangement on fixed structures. Analysis of synchrony-optimized networks reveals three characteristic correlations---positive frequency--out-degree, negative neighbor frequency, and negative mirror-node frequency—that together shape a coherent picture of how directed topology and frequency distributions co-organize to support synchronization. These findings offer practical strategies for stabilizing power grids, regulating neural rhythms, and designing resilient communication systems, while opening avenues toward higher-order interactions and exotic collective states such as explosive synchronization and chimera states.
\end{quotation}


\section{Introduction}
Multilayer network frameworks have emerged as powerful mathematical tools for representing complex systems composed of interacting components across multiple contexts~\cite{boccaletti2014structure,bianconi2018multilayer}. While the traditional single-network approach has been highly successful in explaining and predicting the behavior of complex systems~\cite{albert2002statistical,newman2003structure,boccaletti2006complex}, it often overlooks the fact that real-world entities typically engage in more than one type of interaction simultaneously. By incorporating multiple interconnected layers, multilayer networks provide a more realistic and comprehensive representation, where each layer captures distinct types of relationships, such as social interactions occurring across platforms like Facebook and Twitter~\cite{aleta2019multilayer}. This enriched framework has attracted significant attention in recent years, leading to the development of theoretical approaches for analyzing interconnected network layers with diverse structural and functional characteristics. Within this setting, a broad spectrum of dynamical processes has been investigated, including information and innovation diffusion~\cite{myers2012information}, swarm coordination and control~\cite{vicsek1995novel,prorok2017impact}, brain connectivity~\cite{vaiana2020multilayer,lim2019discordant,battiston2017multilayer}, epidemic spreading~\cite{wei2016cooperative}, cascading dynamics~\cite{brummitt2012suppressing}, and evolutionary processes~\cite{gomez2012evolution,perc2013evolutionary}.

Despite these advances, a large portion of the existing literature primarily focuses on undirected multilayer structures, where both intra-layer and inter-layer interactions are assumed to be bidirectional. This assumption, however, does not accurately reflect the nature of many real-world systems, which are inherently directional. Prominent examples include the world-wide web~\cite{tadic2001dynamics}, where hyperlinks point from one page to another, food webs~\cite{mackay2020directed} characterized by energy transfer from prey to predator, communication networks~\cite{chen2013identifying}, metabolic and gene regulatory networks~\cite{segal2003module}, citation networks~\cite{newman2004coauthorship}, and power grids and transportation networks in infrastructure systems~\cite{menck2014dead,banavar1999size}. Furthermore, many dynamical processes exhibit intrinsic directionality regardless of whether the underlying connectivity is undirected---for instance, the flow of current or information typically proceeds from higher to lower potential, and infectious diseases spread from infected to susceptible individuals. In addition, systems with hierarchical or multi-level interactions often involve directed dependencies, as seen in gene regulatory mechanisms that capture both protein-DNA and protein-protein interactions~\cite{hecker2009gene}. Consequently, incorporating directionality into multilayer network models is essential for achieving a more faithful and insightful representation of complex systems with structured, multi-level, and asymmetric interactions.

Building on the importance of multilayer representations, it is natural to focus on the dynamical processes that unfold on such structures. Among these, synchronization~\cite{boccaletti2002synchronization,pikovsky2003universal,boccaletti2018synchronization,wu2024synchronization} has emerged as one of the most extensively studied and fundamentally important collective phenomena in complex networks. It is ubiquitously observed in a wide variety of natural and engineered systems, ranging from the coordinated flashing of fireflies~\cite{buck1988synchronous} and rhythmic applause in large gatherings~\cite{neda2000sound} to neural activity in the brain~\cite{penn2016network}, oscillatory behavior in cellular populations~\cite{de2007dynamical}, and engineered systems such as coupled mechanical oscillators, power grid systems, and large-scale infrastructure networks~\cite{motter2013spontaneous}. To investigate these mechanisms, a wide class of models based on interacting nonlinear dynamical units has been developed. In the absence of interactions, these units typically evolve independently, leading to incoherent dynamics. However, when coupled through an underlying network structure, they can exhibit a collective transition from disorder to coherence, resulting in synchronized behavior. A seminal framework for studying this phenomenon was introduced by Yoshiki Kuramoto~\cite{kuramoto1984chemical}, who proposed a phase oscillator model in which each oscillator is characterized by its own intrinsic frequency and interacts with others through a periodic coupling function. Depending on the interplay between network topology and dynamical parameters, the transition to synchronization in such systems can occur in a continuous, discontinuous, or even explosive manner~\cite{gomez2007paths,gomez2011explosive,zhang2015explosive,kumar2020interlayer,wu2022double,pal2025explosive}. Subsequent extensions of this framework, notably by Sakaguchi and Kuramoto~\cite{sakaguchi1986soluble}, introduced a phase-lag parameter that accounts for delays or frustration in the interactions among oscillators. This modification has proven essential for modeling a wide range of real-world systems, including Josephson junction arrays~\cite{wiesenfeld1998frequency}, power grid dynamics~\cite{dorfler2012synchronization}, and neuronal networks. The inclusion of such a phase-lag term enriches the system dynamics, giving rise to complex behaviors such as chimera states~\cite{abrams2004chimera} and other nontrivial synchronization patterns~\cite{kundu2017transition,khanra2018explosive,kumar2021explosive,seif2025double}. In particular, it alters the effective collective frequency and strongly influences the onset and nature of synchronization transitions. In many cases, this frustration in the coupling can hinder or even suppress the emergence of global synchrony, making the study of synchronization in such settings both challenging and practically significant.

Following this, it is crucial to recognize that the emergence of synchronization is not governed solely by the intrinsic dynamics of the oscillators, but is also strongly influenced by the underlying network structure~\cite{chen2022searching}. In particular, structural features such as degree heterogeneity, coupling asymmetry, and correlations between nodal properties and topology play a decisive role in shaping collective dynamics. This interplay has motivated extensive efforts toward optimizing synchronization through suitable tuning of frequencies and network architecture. Early studies revealed that enhanced synchrony can be achieved when neighboring oscillators exhibit negatively correlated frequencies and when higher-degree nodes are associated with larger frequency magnitudes~\cite{brede2008synchrony}. Building on this idea, alignment-based frameworks were later introduced to systematically relate oscillator dynamics with network structure, enabling optimization either through appropriate frequency assignment or structural rewiring~\cite{skardal2014optimal}. Alternative approaches based on dimension reduction further provided analytical insight into the collective dynamics and offered strategies for network design, although they primarily capture limited correlation patterns and are not sufficient for achieving maximal synchronization~\cite{ott2008low,gottwald2015model,pinto2015optimal}.

More recent advances have extended these optimization principles to directed networks, where the roles of in-degree and out-degree become fundamentally distinct, and strong correlations between nodal dynamics and incoming connectivity are shown to significantly promote synchronization~\cite{skardal2016optimal}. In parallel, analytical approaches have been developed to determine optimal frequency configurations that ensure perfect synchronization even in the presence of phase frustration, with robustness observed in the vicinity of such optimal states~\cite{kundu2017perfect}. These ideas have also been generalized to multiplex settings, where both structural and dynamical information across layers are incorporated through a multiplex synchrony alignment framework to enhance collective behavior~\cite{kundu2020optimizing}. Recent work has also explored synchronization optimization in directed phase-frustrated networks using alignment-based formulations, along with network reconstruction strategies to sustain coherence beyond optimal conditions~\cite{das2025perfect}. Despite these important developments, a unified framework that simultaneously incorporates directed interactions, multilayer architecture, and frustrated dynamics remains largely unexplored.

In this work, we address these challenges by developing an analytical framework for achieving optimal synchronization in directed multiplex networks of phase oscillators, considering both frustrated and non-frustrated regimes. To this end, we introduce a multiplex synchrony alignment function (MSAF), which integrates the structural properties of individual directed layers with the dynamical characteristics of the oscillators. Using this formulation, we derive two distinct classes of optimal frequency configurations: one that ensures perfect synchronization at a prescribed coupling strength in the presence of phase-lag, and another that maximizes synchronization over a broad range of coupling values. In addition, we explore network-level optimization through a directed link rewiring strategy aimed at minimizing the MSAF, along with a swapping-based method for optimally assigning a fixed set of frequencies across layers. Our analysis, supported by extensive numerical simulations on directed duplex networks, demonstrates that the proposed strategies significantly outperform conventional distributions. Furthermore, the resulting synchrony-optimized configurations reveal several nontrivial structural-dynamical correlations, including a positive relationship between nodal frequencies and out-degree, a negative correlation between neighboring frequencies, and an anti-correlation between mirror node frequencies across layers. These findings provide new insights into the mechanisms governing synchronization in directed multilayer systems and underscore the importance of jointly considering structure, dynamics, and interlayer interactions in the design of optimally synchronized complex networks.


\section{Model}
\begin{figure}
	\centering
	\includegraphics[width=6cm]{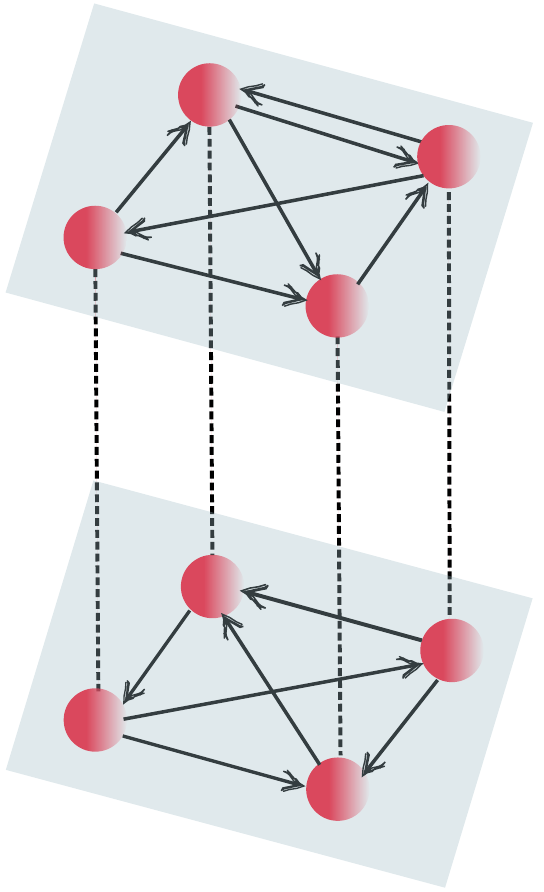}
	\caption{Schematic diagram of a duplex network composed of two directed layers. The dashed lines represent interlayer links between mirror nodes in the two layers.}		
	\label{fig_dir}
\end{figure}

We consider a multiplex system consisting of two directed layers, where the connectivity within each layer is encoded by its own adjacency matrix. The nodes in each layer are modeled as Kuramoto oscillators, capturing their intrinsic dynamic behavior. Mirror nodes---pairs of nodes representing the same entity across different layers---facilitate interlayer coupling, enabling information exchange and influencing the synchronization patterns across the network. A schematic representation of such a duplex network is provided in Fig.~\ref{fig_dir}, where interlayer connections between mirror nodes are indicated by dashed lines.

While the classical Kuramoto model assumes all-to-all interactions~\cite{kuramoto1984chemical}, subsequent extensions have adapted it to more complex and realistic network topologies~\cite{arenas2008synchronization}. In this work, we employ a generalized Sakaguchi–Kuramoto model~\cite{sakaguchi1986soluble} for duplex networks with directed intra-layer connections, incorporating phase-lag parameters to account for frustration or time-delay effects in the interactions. The phase evolution of the $i$-th node in layer 1 and its mirror in layer 2 is governed by the following coupled equations:

\begin{equation}\label{eqn1}
	\scalebox{0.86}{$
		\begin{aligned}
			\frac{d\theta_{i}^{(1)}}{dt}=\omega_{i}^{(1)}+K \sum_{j=1}^{N} A_{ij}^{(1)} \sin\left(\theta_{j}^{(1)}-\theta_{i}^{(1)}-\alpha^{(1)}\right)+ K \sin\left(\theta_{i}^{(2)}-\theta_{i}^{(1)}\right), \\
			\frac{d\theta_{i}^{(2)}}{dt}=\omega_{i}^{(2)}+K \sum_{j=1}^{N} A_{ij}^{(2)} \sin\left(\theta_{j}^{(2)}-\theta_{i}^{(2)}-\alpha^{(2)}\right)+ K \sin\left(\theta_{i}^{(1)}-\theta_{i}^{(2)}\right), \\
		\end{aligned}    
		$}
\end{equation}
where $i=1,2,\dots,N$; $\theta_{i}^{(l)}$ denotes the phase of the $i$-th node in layer $l~(l=1,2)$, and $\omega_{i}^{(l)}$ represents its natural frequency. The coupling strength $K$ is taken to be the same for all interactions and across both layers, ensuring uniformity in the coupling throughout the duplex network. Each layer consists of $N$ nodes. The directed connectivity within layer $l$ is captured by the adjacency matrix $A_{ij}^{(l)}$, with $A_{ij}^{(l)}=1$ if there exists a directed link from node $i$ to node $j$, and $0$ otherwise. Frustration parameters $\alpha^{(l)}$ are incorporated into the intralayer coupling terms to account for layer-specific phase lags, thereby influencing the local synchronization dynamics within each layer.

To quantify the level of synchronization in each layer, we introduce the Kuramoto order parameter given by
\begin{equation}
	R_le^{\iota\psi_{l}}=\frac{1}{N}\sum_{j=1}^{N} e^{\iota\theta_{j}^{(l)}}. \label{eqn2}
\end{equation}

The synchronization order parameter $R_l$ takes values in the interval $[0,1]$. A value of $R_l=1$ corresponds to full phase synchronization within the layer, meaning all oscillators share the same phase. Conversely, $R_l=0$ indicates a completely asynchronous state. The quantity $\psi_{l}$ denotes the mean phase of the oscillators in layer $l$ and lies in the range $[0,2\pi)$.

To investigate the collective behavior of the entire duplex network, we define the global order parameter as:
\begin{equation}
	Re^{\iota \psi}=\frac{1}{2N} \sum_{l=1}^{2}\sum_{j=1}^{N}e^{\iota \theta_{j}^{(l)}}, \label{eqn3}
\end{equation}
where $R\in[0,1]$ quantifies the overall phase coherence across all oscillators in the system. The quantity $\psi$ denotes the global collective phase of the duplex network. Building upon the dynamical equations in Eq.~(\ref{eqn1}) alongside the layer-specific and global order parameters defined in Eqs.~(\ref{eqn2}) and (\ref{eqn3}), we develop an analytical approach to investigate optimal and perfect synchronization in multiplex networks of phase oscillators, considering scenarios with and without phase frustration. A comprehensive theoretical framework is provided in the next section.


\section{Theoretical framework}\label{sec:theoretical}
Here we describe an analytical framework to derive two distinct frequency sets following the approach established in Refs.~\cite{skardal2016optimal,das2025perfect}: one that enables optimal synchronization and another that achieves perfect synchronization at a targeted point in the parameter space. First, we derive a multiplex synchronization alignment function (MSAF) for multiplex networks. This function is then utilized to obtain the frequency sets that facilitate optimal and perfect synchronization in the network.

In the coherent regime where phase differences are small $\left(\abs{\theta_{j}^{(l)}-\theta_{i}^{(l)}} \ll 1\right)$, we linearize the system~(\ref{eqn1}) by applying a first-order Taylor expansion. This yields the following approximated dynamics:
\begin{equation}\label{eqn4}
	\begin{aligned}
		\frac{d\theta_{i}^{(1)}}{dt}&=\tilde{\omega}_{i}^{(1)} - K \cos\alpha^{(1)} \sum_{j=1}^{N} L_{ij}^{(1)} \theta_{j}^{(1)} + K \left(\theta_{i}^{(2)}-\theta_{i}^{(1)}\right), \\
		\frac{d\theta_{i}^{(2)}}{dt}&=\tilde{\omega}_{i}^{(2)} - K \cos\alpha^{(2)} \sum_{j=1}^{N} L_{ij}^{(2)} \theta_{j}^{(2)}
		+ K \left(\theta_{i}^{(1)}-\theta_{i}^{(2)}\right),
	\end{aligned}
\end{equation}
where the effective natural frequency $\tilde{\omega}_{i}^{(l)}$ is given by $\tilde{\omega}_{i}^{(l)} = \omega_{i}^{(l)} + K  s_{i}^{\mathrm{out}(l)}\sin(-\alpha^{(l)})$ and the Laplacian matrix for layer $l$ is defined as $L_{ij}^{(l)}=\delta_{ij}s_{i}^{\mathrm{out}(l)} - A_{ij}^{(l)},~(l=1,2)$. Here, $s_{i}^{\mathrm{out}(l)}=\sum_{j=1}^{N} A_{ij}^{(l)}$ denotes the out-degree of node $i$ in layer $l$, and $\delta_{ij}$ is the Kronecker delta.

In vector form, Eq.~(\ref{eqn4}) can be expressed more compactly as
\begin{equation}\label{eqn5}
	\begin{aligned}
		\dot{\pmb{\theta}}^{(1)}&=\tilde{\pmb{\omega}}^{(1)}-K \cos\alpha^{(1)}~ \pmb{L}^{(1)}\pmb{\theta}^{(1)} + K \left(\pmb{\theta}^{(2)}-\pmb{\theta}^{(1)}\right), \\
		\dot{\pmb{\theta}}^{(2)}&=\tilde{\pmb{\omega}}^{(2)}-K \cos\alpha^{(2)}~ \pmb{L}^{(2)}\pmb{\theta}^{(2)} + K \left(\pmb{\theta}^{(1)}-\pmb{\theta}^{(2)}\right),
	\end{aligned}
\end{equation}
where $\pmb{\theta}^{(l)}=\left[\theta_{1}^{(l)},\theta_{2}^{(l)},\dots,\theta_{N}^{(l)} \right]^{\mathrm{T}}$ and $\tilde{\pmb{\omega}}^{(l)}=\left[\tilde{\omega}_{1}^{(l)},\tilde{\omega}_{2}^{(l)},\dots,\tilde{\omega}_{N}^{(l)} \right]^{\mathrm{T}}$ are $N$-dimensional column vectors for layer $l~(l=1,2)$. The matrix $\pmb{L}^{(l)}$ denotes the Laplacian of layer $l$ with entries $L_{ij}^{(l)}$.

Now imposing the steady-state condition $\dot{\pmb{\theta}}^{(1)}=\dot{\pmb{\theta}}^{(2)}=0$ on Eq.~(\ref{eqn5}) leads to the following expressions for the steady-state phases:
\begin{equation}\label{eqn6}
	\begin{aligned}
		{\pmb{\theta}^{(1)}}^{\ast}=\frac{1}{K}{\pmb{L}_{m}^{(1)^\dagger}\tilde{\pmb{\omega}}_{m}^{(1)}}, \\
		{\pmb{\theta}^{(2)}}^{\ast}=\frac{1}{K}{\pmb{L}_{m}^{(2)^\dagger}\tilde{\pmb{\omega}}_{m}^{(2)}},
	\end{aligned}
\end{equation}
where the modified Laplacian matrices $\pmb{L}_{m}^{(1)}$ and $\pmb{L}_{m}^{(2)}$ are defined as
\begin{equation}\label{eqn7}
	\begin{aligned}
		&\pmb{L}_{m}^{(1)}=\cos\alpha^{(1)}\pmb{L}^{(1)}+\cos\alpha^{(2)}\pmb{L}^{(2)} +\cos\alpha^{(1)}\cos\alpha^{(2)}\pmb{L}^{(2)}\pmb{L}^{(1)}, \\
		&\pmb{L}_{m}^{(2)}=\cos\alpha^{(1)}\pmb{L}^{(1)}+\cos\alpha^{(2)} \pmb{L}^{(2)}+\cos\alpha^{(1)}\cos\alpha^{(2)}\pmb{L}^{(1)}\pmb{L}^{(2)},
	\end{aligned}
\end{equation}
and the modified frequency vectors $\tilde{\pmb{\omega}}_{m}^{(1)}$ and $\tilde{\pmb{\omega}}_{m}^{(2)}$ are given by
\begin{equation}\label{eqn8}
	\begin{aligned}
		&\tilde{\pmb{\omega}}_{m}^{(1)}=\tilde{\pmb{\omega}}^{(1)}+\tilde{\pmb{\omega}}^{(2)}+\cos\alpha^{(2)}\pmb{L}^{(2)} \tilde{\pmb{\omega}}^{(1)}, \\
		&\tilde{\pmb{\omega}}_{m}^{(2)}=\tilde{\pmb{\omega}}^{(1)}+\tilde{\pmb{\omega}}^{(2)}+\cos\alpha^{(1)}\pmb{L}^{(1)} \tilde{\pmb{\omega}}^{(2)}.
	\end{aligned}
\end{equation}
Here, the dagger symbol $(.)^\dagger$ denotes the Moore–Penrose pseudoinverse~\cite{ben1974generalized}.

Since $\pmb{L}_{m}^{(l)}$ is a real $N\times N$ matrix for each layer $l~(l=1,2)$, it admits a singular value decomposition (SVD). According to the SVD theorem~\cite{golub1996matrix}, there exist orthogonal matrices $\pmb{U}^{(l)}=\left[\pmb{u}_{1}^{(l)},\pmb{u}_{2}^{(l)},\dots,\pmb{u}_{N}^{(l)}\right]$ and $\pmb{V}^{(l)}=\left[\pmb{v}_{1}^{(l)},\pmb{v}_{2}^{(l)},\dots,\pmb{v}_{N}^{(l)}\right]$ such that ${\left(\pmb{U}^{(l)}\right)}^{\mathrm{T}} \pmb{L}_{m}^{(l)} \pmb{V}^{(l)}=\pmb{\Sigma}^{(l)}=\mathrm{diag}\left(\sigma_{1}^{(l)},\sigma_{2}^{(l)},\dots,\sigma_{N}^{(l)}\right)$, where the singular values satisfy $\sigma_{1}^{(l)}\ge\sigma_{2}^{(l)}\ge \dots \ge\sigma_{N}^{(l)}\ge0$. The vectors $\pmb{u}_{i}^{(l)}$ and $\pmb{v}_{i}^{(l)}$ denote the $i$-th left and right singular vectors of $\pmb{L}_{m}^{(l)}$, respectively. For strongly connected networks in both layers, the singular values can be ordered without loss of generality as $0=\sigma_{1}^{(l)}<\sigma_{2}^{(l)}\le \dots \le\sigma_{N}^{(l)}$. Under this ordering, the matrix $\pmb{L}_{m}^{(l)}$ admits the decomposition $\pmb{L}_{m}^{(l)}=\sum_{j=2}^{N} \sigma_{j}^{(l)} \pmb{u}_{j}^{(l)} \left(\pmb{v}_{j}^{(l)}\right)^{\mathrm{T}}$. Defining the diagonal matrix $\left(\pmb{\Sigma}^{(l)}\right)^\dagger=\mathrm{diag}\left(0,1/\sigma_{2}^{(l)},1/\sigma_{3}^{(l)},\dots,1/\sigma_{N}^{(l)} \right)$, the Moore–Penrose pseudoinverse of $\pmb{L}_{m}^{(l)}$ is then given by
\begin{equation}
	\left(\pmb{L}_{m}^{(l)}\right)^\dagger=\pmb{V}^{(l)} \left(\pmb{\Sigma}^{(l)}\right)^\dagger \left(\pmb{U}^{(l)}\right)^{\mathrm{T}}=\sum_{j=2}^{N} \frac{\pmb{v}_{j}^{(l)} \left(\pmb{u}_{j}^{(l)}\right)^{\mathrm{T}} }{\sigma_{j}^{(l)}}. \label{eqn9}
\end{equation}

The order parameter for the $l$-th layer, defined in Eq.~(\ref{eqn2}), can then be expressed as
\begin{equation}
	R_{l}=1-\frac{1}{2N} \norm{{\pmb{\theta}^{(l)}}^{\ast}}^2, \label{eqn10}
\end{equation}
where the squared norm of the steady-state phase vector is given by
\begin{align*}
	\norm{{\pmb{\theta}^{(l)}}^{\ast}}^2&=\frac{1}{K^2} \innerproduct{{\pmb{L}_{m}^{(l)^\dagger}\tilde{\pmb{\omega}}_{m}^{(l)}},{\pmb{L}_{m}^{(l)^\dagger}\tilde{\pmb{\omega}}_{m}^{(l)}}} \\
	&= \frac{1}{K^2} \sum_{j=2}^{N} \frac{{\innerproduct{\pmb{u}_{j}^{(l)},\tilde{\pmb{\omega}}_{m}^{(l)}}}^2}{{\left(\sigma_{j}^{(l)}\right)}^2}.
\end{align*}
Now substituting the above expression into Eq.~(\ref{eqn10}) yields
\begin{equation}
	R_{l}=1-\frac{1}{2K^2}J\left(\tilde{\pmb{\omega}}_{m}^{(l)},\pmb{L}_{m}^{(l)}\right), \label{eqn11}
\end{equation}
where $J(.,.)$ is the \textit{multiplex synchrony alignment function} (MSAF) for the $l$-th layer defined as
\begin{equation}
	J\left(\tilde{\pmb{\omega}}_{m}^{(l)},\pmb{L}_{m}^{(l)}\right)=\frac{1}{N} \sum_{j=2}^{N} \frac{{\innerproduct{\pmb{u}_{j}^{(l)},\tilde{\pmb{\omega}}_{m}^{(l)}}}^2}{{\left(\sigma_{j}^{(l)}\right)}^2}. \label{eqn12}
\end{equation}
Notably, the MSAF depends on $\pmb{L}_{m}^{(l)}$ and $\tilde{\pmb{\omega}}_{m}^{(l)}$, which themselves incorporate the Laplacians and dynamics of both layers through Eqs.~(\ref{eqn7}) and (\ref{eqn8}). This distinguishes the present formulation from its monolayer counterpart in the absence of frustration, as reported in~\cite{skardal2016optimal}. Eq.~(\ref{eqn11}) reveals that the MSAF plays a crucial role in determining the synchronization level. An increase in the MSAF value corresponds to a deterioration of synchronization, whereas in the limit $J\left(\tilde{\pmb{\omega}}_{m}^{(l)},\pmb{L}_{m}^{(l)}\right) \longrightarrow0$, the order parameter approaches $R_l \longrightarrow1$, indicating enhanced synchronization.

The global order parameter of the system, defined in Eq.~(\ref{eqn3}), takes the form
\begin{align}
	R&=1-\frac{1}{4N} \norm{{\pmb{\theta}^{(1)}}^{\ast}}^2 - \frac{1}{4N} \norm{{\pmb{\theta}^{(2)}}^{\ast}}^2 \nonumber \\
	&=1-\frac{1}{4K^2} J\left(\tilde{\pmb{\omega}}_{m}^{(1)},\pmb{L}_{m}^{(1)}\right) - \frac{1}{4K^2} J\left(\tilde{\pmb{\omega}}_{m}^{(2)},\pmb{L}_{m}^{(2)}\right). \label{eqn13}
\end{align}
From this expression, it is evident that minimizing the MSAF for each layer simultaneously maximizes both the individual layer order parameters and the global order parameter of the entire duplex network.

For perfect synchronization in each layer, the order parameter must satisfy $R_{l}=1$, which from Eq.~(\ref{eqn11}) requires $J\left(\tilde{\pmb{\omega}}_{m}^{(l)},\pmb{L}_{m}^{(l)}\right)=0,~l=1,2$. The simplest choice fulfilling this condition is $\tilde{\pmb{\omega}}_{m}^{(l)}=\pmb{0}$ for $l=1,2$. From Eq.~(\ref{eqn8}), this condition can be satisfied by setting $\tilde{\pmb{\omega}}^{(1)}=\tilde{\pmb{\omega}}^{(2)}=\pmb{0}$ simultaneously. Consequently, the natural frequencies for each layer that yield perfect synchronization are given by
\begin{equation}\label{eqn14}
	\begin{aligned}
		\pmb{\omega}^{(1)}&=K_{\mathrm{p}} \pmb{s}^{\mathrm{out}(1)} \sin\alpha^{(1)}, \\
		\pmb{\omega}^{(2)}&=K_{\mathrm{p}} \pmb{s}^{\mathrm{out}(2)} \sin\alpha^{(2)},
	\end{aligned}
\end{equation}
where $K_{\mathrm{p}}$ denotes the targeted coupling strength at which perfect synchronization is achieved in the multiplex network for given non-zero frustration parameters $\alpha^{(1)}$ and $\alpha^{(2)}$.

To avoid the trivial solution, we consider natural frequency vectors with zero mean, $\innerproduct{\pmb{\mathrm{x}}}=\frac{1}{N} \sum_{j=1}^{N} x_{j}=0$, and fixed nonzero standard deviation, $\sigma^2=\frac{1}{N} \sum_{j=1}^{N} \left(x_{j}-\innerproduct{\pmb{\mathrm{x}}} \right)^2>0$. Expanding $\tilde{\pmb{\omega}}_{m}^{(l)}$ in the basis of left singular vectors as $\tilde{\pmb{\omega}}_{m}^{(l)}=\sum_{j=1}^{N} a_{j}^{(l)} \pmb{u}_{j}^{(l)}$ and substituting into the expression for the MSAF in Eq.~(\ref{eqn12}), we find that $J\left(\tilde{\pmb{\omega}}_{m}^{(l)},\pmb{L}_{m}^{(l)}\right)$ can be minimized by concentrating as much weight as possible onto the coefficient $a_{N}^{(l)}$ corresponding to the dominant left singular vector $\pmb{u}_{N}^{(l)}$. However, since the left singular vectors $\pmb{u}_{j}^{(l)}$ generally possess nonzero mean, an additional shift is required to enforce the zero-mean constraint. Following this approach, the optimal choice of $\tilde{\pmb{\omega}}_{m}^{(l)}$ that minimizes the MSAF is given by
\begin{equation}
	\tilde{\pmb{\omega}}_{m}^{(l)}=\pm \sigma\sqrt{N} \left(\frac{\pmb{u}_{N}^{(l)}-\frac{\innerproduct{\pmb{u}_{N}^{(l)}}}{\innerproduct{\pmb{u}_{1}^{(l)}}} \pmb{u}_{1}^{(l)} }{\sqrt{1+\frac{\innerproduct{\pmb{u}_{N}^{(l)}}^2}{\innerproduct{\pmb{u}_{1}^{(l)}}^2} }} \right), \label{eqn15}
\end{equation}
where $\innerproduct{\pmb{u}_{j}^{(l)}}=\frac{1}{N} \sum_{k=1}^{N} {\left(\pmb{u}_{j}^{(l)} \right)}_{k}$ denotes the mean entry of the $j$-th left singular vector of $\pmb{L}_{m}^{(l)}$.

Following the above approach and after straightforward algebraic manipulation, the optimal natural frequencies that maximize synchronization in each layer are obtained as
\begin{equation}\label{eqn16}
	\begin{aligned}
		\pmb{\omega}^{(1)}&={\pmb{L}_{m}^{(2)}}^\dagger \left[\tilde{\pmb{\omega}}_{m}^{(1)} - \tilde{\pmb{\omega}}_{m}^{(2)} + \cos\alpha^{(1)} \pmb{L}^{(1)} \tilde{\pmb{\omega}}_{m}^{(1)} \right] + K\pmb{s}^{\mathrm{out}(1)}\sin\alpha^{(1)}, \\
		\pmb{\omega}^{(2)}&={\pmb{L}_{m}^{(1)}}^\dagger \left[\tilde{\pmb{\omega}}_{m}^{(2)} - \tilde{\pmb{\omega}}_{m}^{(1)} + \cos\alpha^{(2)} \pmb{L}^{(2)} \tilde{\pmb{\omega}}_{m}^{(2)} \right] + K\pmb{s}^{\mathrm{out}(2)}\sin\alpha^{(2)},
	\end{aligned}
\end{equation}
where $\tilde{\pmb{\omega}}_{m}^{(1)}$ and $\tilde{\pmb{\omega}}_{m}^{(2)}$ are the frequency vectors given by Eq.~(\ref{eqn15}). In the non-frustrated case $(\alpha^{(1)}=\alpha^{(2)}=0)$, the frequencies in Eq.~(\ref{eqn16}) reduce to a form independent of the coupling strength $K$. However, in the presence of frustration $(\alpha^{(l)}\neq0)$, the optimal frequencies exhibit an explicit dependence on $K$. This $K$-dependence necessitates an optimal coupling strength $K_{\mathrm{opt}}$ at which the network achieves the best possible synchronization for the chosen frequency set.


\section{Numerical results}
To validate the analytical framework developed in the previous section, we now present numerical simulations of the directed multiplex system (\ref{eqn1}). We begin by examining the non-frustrated case $(\alpha^{(1)}=\alpha^{(2)}=0)$, where the optimal frequencies derived in Eq.~(\ref{eqn16}) are expected to enhance synchronization. Subsequently, we investigate the frustrated regime $(\alpha^{(l)}\neq0)$ and verify the performance of both the perfect frequency set from Eq.~(\ref{eqn14}) and the optimal frequency set from Eq.~(\ref{eqn16}).

\begin{figure*}[ht]
	\centering
	\includegraphics[width=16cm]{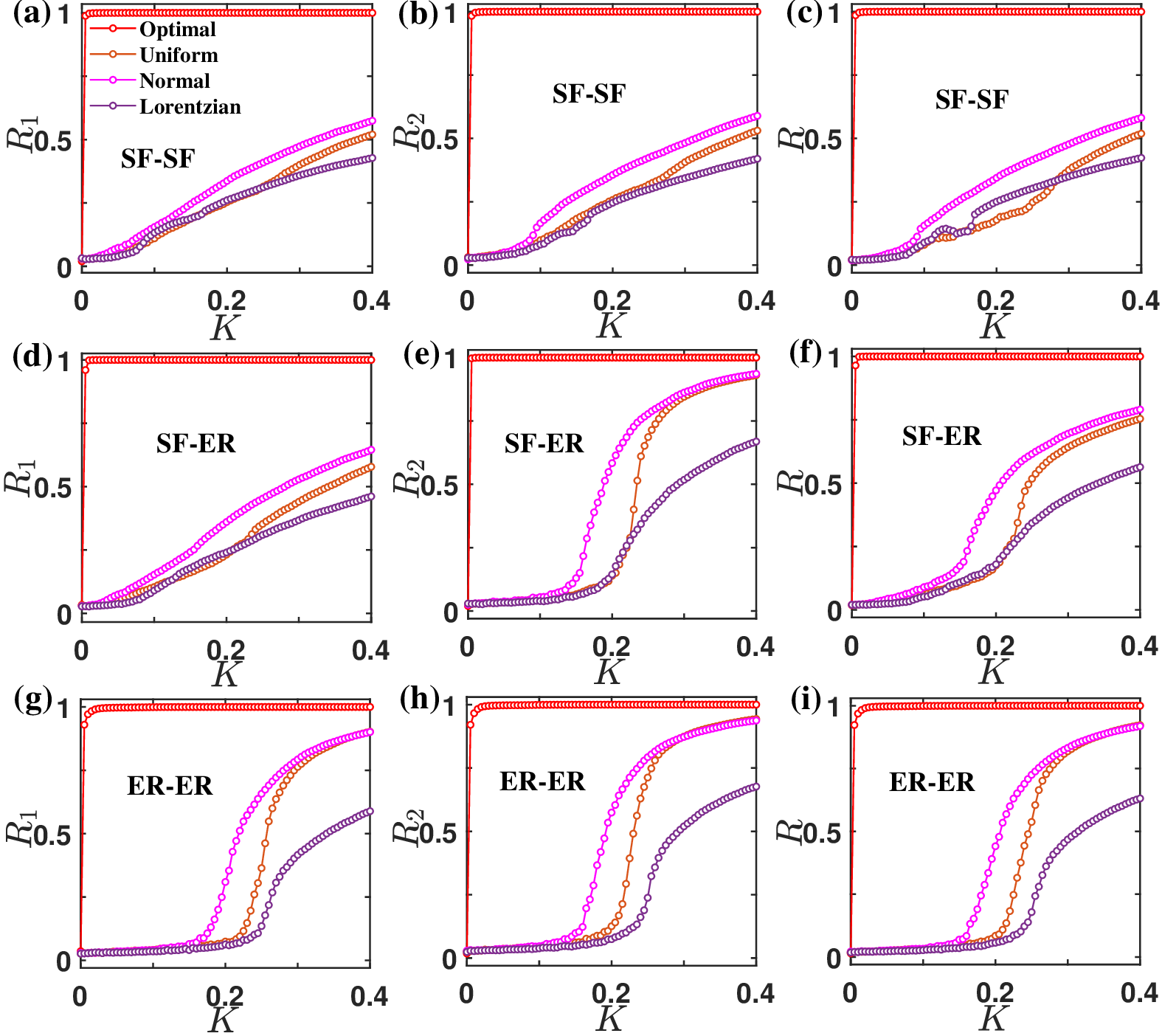}
	\caption{Optimal synchronization in a directed multiplex network of non-frustrated phase oscillators for three duplex configurations: SF–SF (a–c), SF–ER (d–f), and ER–ER (g–i). Panels (a), (d), and (g) show the variation of the order parameter $R_{1}$ of the first layer with coupling strength $K$; panels (b), (e), and (h) display the order parameter $R_{2}$ of the second layer; and panels (c), (f), and (i) depict the global order parameter $R$ of the entire duplex network. The red curves with open circles represent the order parameters obtained using the optimal frequency sets derived in Eq.~(\ref{eqn16}). For comparison, the orange, magenta, and purple curves with open circles correspond to the order parameters obtained using uniform, normal, and Lorentzian distributions of the natural frequencies $\pmb{\omega}^{(l)}$, respectively.
	}
	\label{fig1}
\end{figure*}

For our numerical experiments, we consider a heterogeneous multiplex network comprising two directed layers, each constructed as either a scale-free (SF) network~\cite{bollobas2003directed} or an Erd{\H{o}}s-R{\'e}nyi (ER) network~\cite{erdos1959random,gilbert1959random}. This yields three distinct duplex configurations: SF–SF, SF–ER, and ER–ER. In the SF–SF duplex, layer 1 has an out-degree exponent $\gamma^{\mathrm{out}}=2.56$, while layer 2 has $\gamma^{\mathrm{out}}=2.43$. For the SF–ER configuration, layer 1 is scale-free with the same parameters as above, and layer 2 is an ER graph with connection probability $p=0.0095$. In the ER–ER duplex, layer 1 has connection probability $p=0.0075$, while layer 2 has $p=0.0095$. All layers are directed and strongly connected, with each containing $N=1000$ nodes. These network parameters are employed throughout the numerical study unless stated otherwise.

We then integrate the governing equations using a fourth-order Runge–Kutta (RK4) scheme and compute the layer-specific order parameters $R_l$ from Eq.~(\ref{eqn2}) as well as the global order parameter $R$ from Eq.~(\ref{eqn3}) as functions of the coupling strength $K$.

\begin{figure*}[ht]
	\centering
	\includegraphics[width=\linewidth]{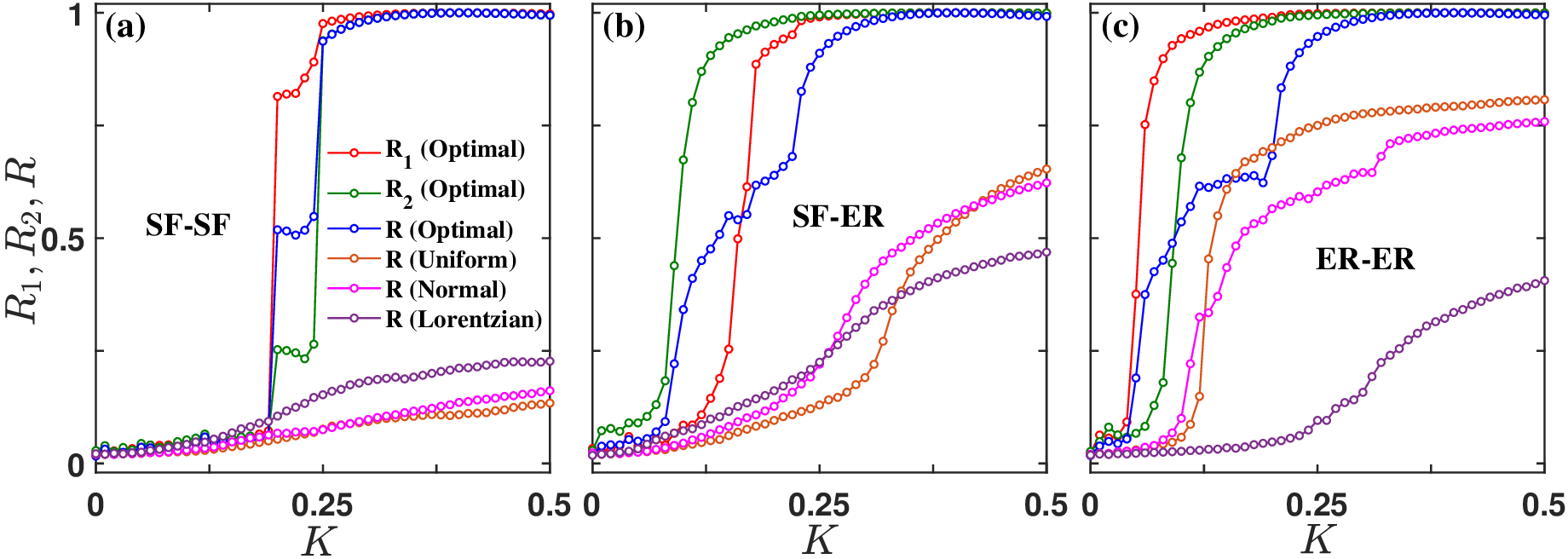}
	\caption{Optimal synchronization in directed frustrated multiplex networks with $\alpha^{(1)}=0.2$ and $\alpha^{(2)}=0.4$ for three duplex configurations: SF–SF (a), SF–ER (b), and ER–ER (c). In each panel, the red, green, and blue curves with open circles denote the layer-specific order parameters $R_1$ and $R_2$, and the global order parameter $R$, respectively, obtained using the optimal frequency sets from Eq.~(\ref{eqn16}). The orange, magenta, and purple curves with open circles show the global order parameter for uniform, normal, and Lorentzian frequency distributions, respectively, serving as a benchmark for comparison.}
	\label{fig2}
\end{figure*}

\subsection{Optimal synchronization in directed non-frustrated duplex networks}
We first consider the non-frustrated regime where the phase-lag parameters vanish, i.e., $\alpha^{(1)}=\alpha^{(2)}=0$. Our aim is to achieve optimal synchronization using the analytically derived frequency sets given in Eq.~(\ref{eqn16}). To this end, we numerically integrate the governing equations in Eq.~(1) for the three directed duplex configurations described above by taking $\sigma=1$. We have also verified the robustness of our results by varying $\sigma$ and consistently observed that the chosen frequency sets facilitate optimal synchronization across the multiplex network. Figure~\ref{fig1} displays the numerically computed order parameters as functions of the coupling strength $K$. Panels (a), (d), and (g) show the order parameter $R_1$ of the first layer; panels (b), (e), and (h) show $R_2$ of the second layer; and panels (c), (f), and (i) show the global order parameter $R$ of the entire duplex network. The three columns correspond to the three network configurations: SF–SF in (a)–(c), SF–ER in (d)–(f), and ER–ER in (g)–(i). For all configurations, the order parameters obtained using the optimal frequency sets from Eq.~(\ref{eqn16}) are plotted as red curves with open circles. For comparison, we also simulate the system using three standard frequency distributions---uniform, normal, and Lorentzian---with the results shown in orange, magenta, and purple curves with open circles, respectively. In all cases, the optimal frequency sets enable the network to achieve synchronization at significantly lower coupling strengths compared to the standard distributions. In contrast, the uniform, normal, and Lorentzian distributions require substantially higher coupling strengths to reach a comparable synchronized state. These results demonstrate that the analytically derived frequency sets effectively enhance synchronization across all three order parameters $R_1$, $R_2$, and $R$ in directed non-frustrated duplex networks.

Having established the effectiveness of the optimal frequency sets in the absence of frustration, we next examine the frustrated regime. In the following subsection, we validate both the optimal frequency sets derived in Eq.~(\ref{eqn16}) and the perfect frequency sets obtained in Eq.~(\ref{eqn14}) for directed multiplex networks with nonzero phase frustration.

\begin{figure*}[ht]
	\centering
	\includegraphics[width=\linewidth]{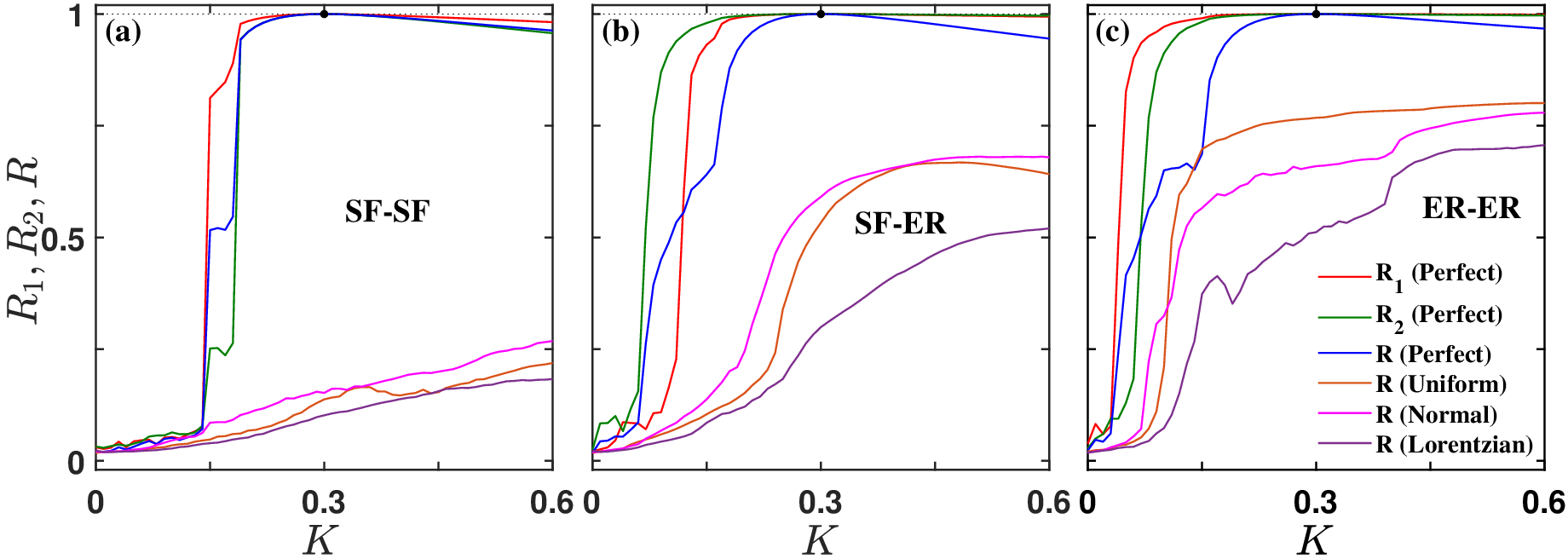}
	\caption{Perfect synchronization in directed frustrated multiplex networks with $\alpha^{(1)}=0.2$ and $\alpha^{(2)}=0.4$, targeted at $K_{\mathrm{p}}=0.3$, for three duplex configurations: SF–SF (a), SF–ER (b), and ER–ER (c). In each panel, the red, green, and blue curves denote the layer-specific order parameters $R_1$ and $R_2$, and the global order parameter $R$, respectively, obtained using the perfect frequency sets from Eq.~(\ref{eqn14}). The orange, magenta, and purple curves correspond to the global order parameter $R$ for uniform, normal, and Lorentzian frequency distributions, respectively.}
	\label{fig3}
\end{figure*}

\begin{figure*}[ht]
	\centering
	\includegraphics[width=\linewidth]{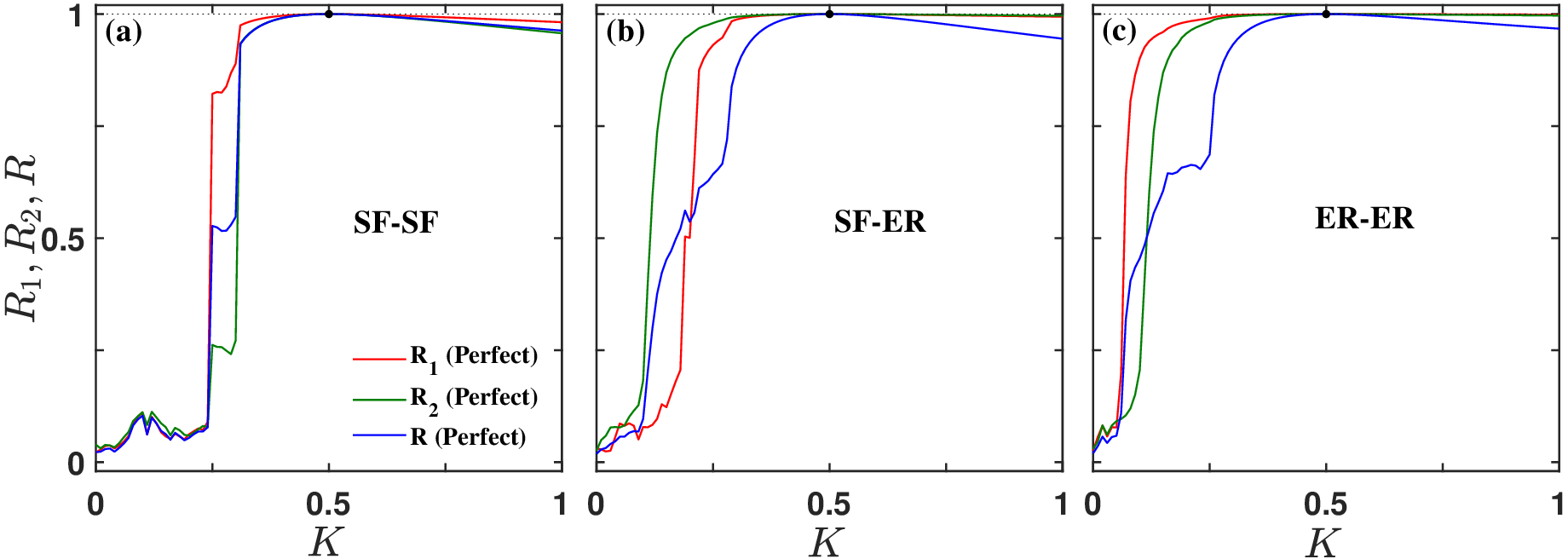}
	\caption{Perfect synchronization in directed frustrated multiplex networks $(\alpha^{(1)}=0.2,\alpha^{(2)}=0.4)$ at the targeted coupling strength $K_{\mathrm{p}}=0.5$ for three duplex configurations: SF–SF (a), SF–ER (b), and ER–ER (c). Red, green, and blue curves denote $R_1$, $R_2$, and $R$, respectively, obtained using the perfect frequency sets from Eq.~(\ref{eqn14}).}
	\label{fig4}
\end{figure*}

\subsection{Synchronization in directed frustrated duplex networks}
We now turn to the frustrated regime where the phase-lag parameters are nonzero, i.e., $\alpha^{(1)}\neq0$ and $\alpha^{(2)}\neq0$. In this case, the last terms in Eq.~(\ref{eqn16}) contain an explicit dependence on the coupling strength $K$ due to the presence of frustration. Consequently, the optimal frequency sets are computed around a desired coupling strength, which we choose as $K_{\mathrm{opt}}=0.4$ for $\alpha^{(1)}=0.2$ and $\alpha^{(2)}=0.4$. The system~(\ref{eqn1}) is then numerically simulated using these optimal frequencies. For other choices of $K_{\mathrm{opt}}$, analogous optimal synchronization behavior is observed. Figure~\ref{fig2} presents the synchronization results for the optimal frequency sets in the frustrated regime. Panels (a), (b), and (c) correspond to the SF–SF, SF–ER, and ER–ER duplex configurations, respectively. In each panel, the red, green, and blue curves with open circles denote the order parameters $R_1$, $R_2$, and the global order parameter $R$, respectively, obtained using the optimal frequency sets from Eq.~(\ref{eqn16}). For comparison, the orange, magenta, and purple curves with open circles represent the global order parameter obtained using uniform, normal, and Lorentzian frequency distributions. From the figure, it is evident that the system reaches a synchronized state at $K=0.4$ when the optimal frequency sets are employed, whereas for the other frequency distributions, the system remains far from synchronization.

We next validate the perfect frequency sets derived in Eq.~(\ref{eqn14}). These sets are designed to achieve perfect synchronization at targeted coupling strengths $K_{\mathrm{p}}$ for given frustration parameters. Figure~\ref{fig3} shows the results for $K_{\mathrm{p}}=3$ with $\alpha^{(1)}=0.2$ and $\alpha^{(2)}=0.4$. Panels (a)–(c) display the order parameters for the SF–SF, SF–ER, and ER–ER configurations, respectively. The red, green, and blue curves represent $R_1$, $R_2$, and the global order parameter $R$ obtained using the perfect frequency sets. For comparison, the orange, magenta, and purple curves show the global order parameter for uniform, normal, and Lorentzian frequency distributions. At the targeted coupling strength $K_{\mathrm{p}}=0.3$, the perfect frequency sets yield $R_1=1$, $R_2=1$ and consequently $R=1$, indicating perfect synchronization across both layers, whereas the standard distributions fail to achieve the same level of coherence.

To further demonstrate the versatility of the perfect frequency sets, we consider another targeted coupling strength $K_{\mathrm{p}}=0.5$ while keeping the frustration parameters unchanged at $\alpha^{(1)}=0.2$ and $\alpha^{(2)}=0.4$. The corresponding results are presented in Fig.~\ref{fig4} for the three duplex configurations. In each panel, the red, green, and blue curves denote $R_1$, $R_2$, and $R$ obtained using the perfect frequency sets from Eq.~(\ref{eqn14}). At $K_{\mathrm{p}}=0.5$, the system again achieves perfect synchronization, confirming that the proposed frequency sets successfully facilitate perfect synchronization at the prescribed coupling strength. These numerical results collectively validate the analytical framework developed in Sec.~\ref{sec:theoretical}, demonstrating that both the optimal and perfect frequency sets effectively enhance synchronization in directed frustrated duplex networks.

Having established that the analytically derived frequency sets significantly enhance synchronization in directed duplex networks, we now consider the inverse problem: given a fixed set of natural frequencies, can we design the network topology to optimize synchronization? In the following section, we address this question by developing a rewiring-based optimization framework that minimizes the multiplex synchronization alignment function (MSAF) for both layers simultaneously, thereby enabling the construction of network structures tailored for optimal synchronization.


\section{Designing optimal duplex networks}\label{design-networks}

\begin{figure*}[ht]
	\centering
	\includegraphics[width=\linewidth]{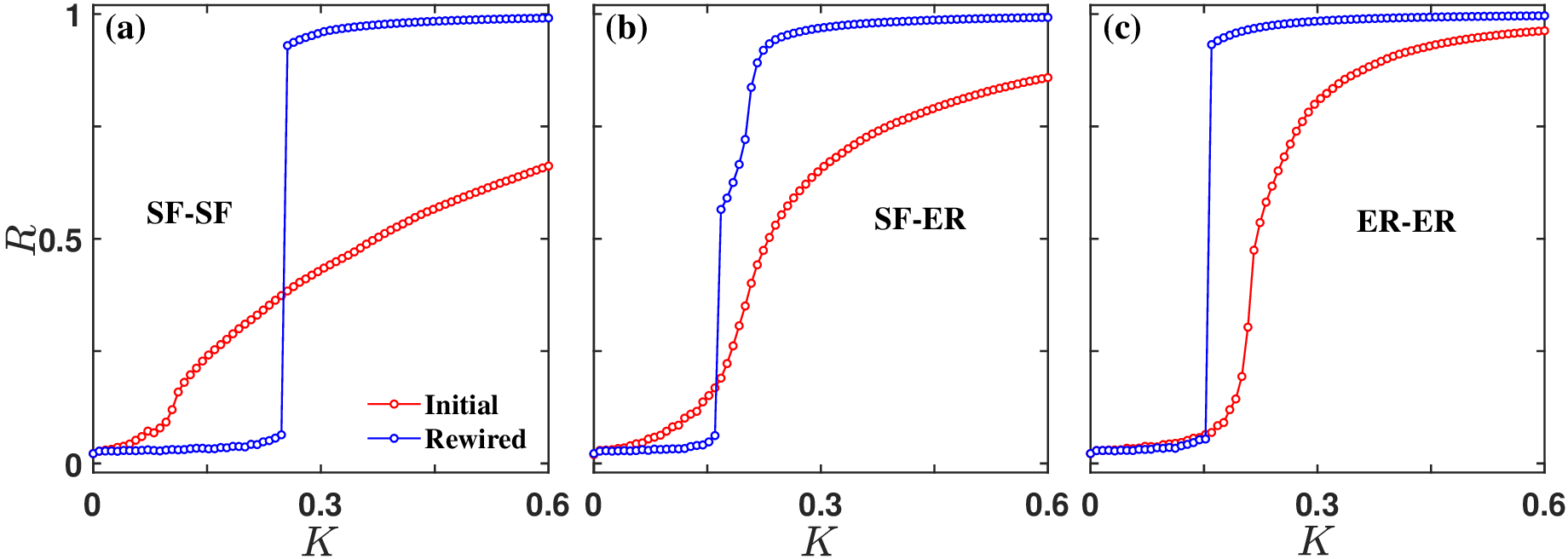}
	\caption{Optimal network design through simultaneous rewiring of both layers to enhance global synchronization. The global order parameter $R$ is plotted as a function of the coupling strength $K$ for the initial (red curves with open circles) and rewired (blue curves with open circles) networks. Panels (a), (b), and (c) correspond to the SF–SF, SF–ER, and ER–ER duplex configurations, respectively. The natural frequencies are drawn from a normal distribution.}
	\label{fig5}
\end{figure*}

We now consider the inverse problem of network construction. We assume that a set of pre-chosen natural frequencies is given, and a duplex network must be built to best synchronize the oscillators. Both layers are taken to be strongly connected, and the natural frequencies $\pmb{\omega}^{(l)}$ for $l=1,2$ are fixed. For a given initial network, we compute the multiplex synchronization alignment function (MSAF) $J\left(\tilde{\pmb{\omega}}_{m}^{(l)},\pmb{L}_{m}^{(l)}\right)$ for each layer, where $\tilde{\pmb{\omega}}_{m}^{(l)}$ and $\pmb{L}_{m}^{(l)}$ depend on the natural frequencies and the Laplacian matrices of both layers through Eqs.~(\ref{eqn7}) and (\ref{eqn8}). Since the MSAF is influenced by the topology of both layers, we can rewire both layers simultaneously to achieve improved synchronization. We employ an accept–reject rewiring mechanism~\cite{skardal2016optimal} to optimize the network structure. Starting from an initial duplex network, we randomly select a directed link $i\longrightarrow j$ in each layer for deletion and introduce a new link $i'\longrightarrow j'$ between two previously disconnected nodes, chosen uniformly at random. This yields a new duplex network with updated MSAF values $J_{1}\left(\tilde{\pmb{\omega}}_{m}^{(l)},\pmb{L}_{m}^{(l)}\right)$ for $l=1,2$. If the new MSAF is lower than the previous one for both layers, i.e., $J_{1}\left(\tilde{\pmb{\omega}}_{m}^{(l)},\pmb{L}_{m}^{(l)}\right)<J\left(\tilde{\pmb{\omega}}_{m}^{(l)},\pmb{L}_{m}^{(l)}\right)$ for $l=1$ and $2$, the rewiring is accepted; otherwise, it is rejected. This process is repeated until the MSAF values become smaller than a desired threshold for both layers, yielding an optimal duplex network.

To demonstrate the effectiveness of this method, we simulate the dynamics in Eq.~(\ref{eqn1}) on the initial and optimized duplex networks. We initialize the rewiring algorithm with the three duplex configurations: SF–SF, SF–ER, and ER–ER. The natural frequencies $\pmb{\omega}^{(l)}$ for both layers are drawn from a normal distribution and remain fixed throughout the optimization. We have taken $\alpha^{(1)}=\alpha^{(2)}=0$. The global order parameter $R$ obtained from simulations on the initial networks is shown in Fig.~\ref{fig5} with blue curves and open circles, while the results for the optimized networks are shown with red curves and open circles. Panels (a), (b), and (c) correspond to the SF–SF, SF–ER, and ER–ER configurations, respectively. A substantial improvement in synchronization is clearly visible across all configurations. The rewired networks exhibit enhanced synchronization properties, characterized by abrupt transitions from incoherence to strong synchronization compared to the initial networks. One can obtain similar improvements in synchronization for the frustrated case as well.

The results presented above demonstrate that targeted rewiring of both layers can substantially enhance synchronization in directed duplex networks when the natural frequencies are fixed. This establishes that network topology plays a crucial role in shaping the collective dynamics. However, in many practical scenarios, the network structure is predetermined, and the available degrees of freedom lie in the assignment of natural frequencies to the nodes. This complementary perspective motivates the problem addressed in the following section.


\section{Optimal frequency arrangement on fixed duplex networks}\label{freq-arrange}

\begin{figure*}[ht]
	\centering
	\includegraphics[width=\linewidth]{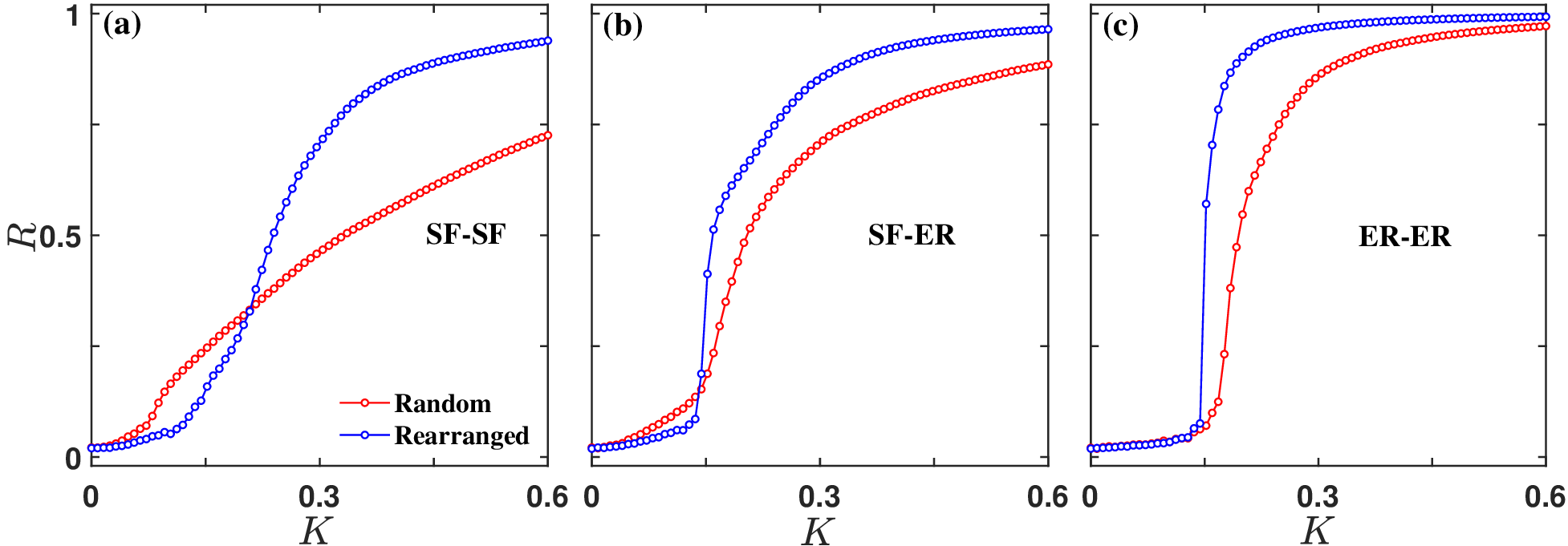}
	\caption{Optimal frequency arrangement on fixed duplex networks to enhance global synchronization. The global order parameter $R$ is plotted as a function of the coupling strength $K$ for random allocation (red curves with open circles) and rearranged allocation obtained through the greedy swapping algorithm (blue curves with open circles). Panels (a), (b), and (c) correspond to the SF–SF, SF–ER, and ER–ER duplex configurations, respectively.}
	\label{fig6}
\end{figure*}

We now consider the problem of oscillator arrangement in both layers of a directed duplex network. Here, we assume that the topology of the duplex network is prescribed, and instead of selecting natural frequencies arbitrarily, we are provided with a fixed set of pre-chosen natural frequencies $\omega^{(l)}$ for $l=1,2$ that must be assigned to the nodes in each layer. Given the combinatorial complexity of this problem, with $N!$ possible arrangements per layer, an exhaustive search is infeasible even for moderately sized networks. We therefore propose a simple greedy algorithm to approximate the optimal arrangement. Initially, the set of natural frequencies---taken here from the standard normal distribution---is assigned randomly to the nodes in both layers. At each step, we propose swapping the frequencies of a randomly selected pair of oscillators in each layer, i.e., $\omega_{i}^{(l)}\longleftrightarrow \omega_{j}^{(l)}$ for $l=1,2$, yielding new frequency vectors $\omega'^{(l)}$. We then compute the updated multiplex synchronization alignment function (MSAF) $J_{1}\left(\tilde{\pmb{\omega}}_{m}^{(l)},\pmb{L}_{m}^{(l)}\right)$ for each layer. If the new MSAF is lower than the previous one for both layers, i.e., $J_{1}\left(\tilde{\pmb{\omega}}_{m}^{(l)},\pmb{L}_{m}^{(l)}\right)<J\left(\tilde{\pmb{\omega}}_{m}^{(l)},\pmb{L}_{m}^{(l)}\right)$ for $l=1$ and $2$, the swap is accepted; otherwise, it is rejected. This process is repeated until the MSAF values become smaller than a desired threshold for both layers, yielding an arrangement that promotes global synchronization. Importantly, the swapping is performed simultaneously in both layers to account for the interdependence introduced by interlayer coupling.

To demonstrate the effectiveness of this approach, we simulate the dynamics in Eq.~(\ref{eqn1}) on the three duplex configurations---SF–SF, SF–ER, and ER–ER---before and after rearranging the natural frequencies using the greedy algorithm. We have taken $\alpha^{(1)}=\alpha^{(2)}=0$. Initially, the frequencies are allocated randomly to the oscillators. The resulting synchronization profiles, showing the global order parameter $R$ as a function of the coupling strength $K$, are presented in Fig.~\ref{fig6} with red curves and open circles. Panels (a), (b), and (c) correspond to the SF–SF, SF–ER, and ER–ER configurations, respectively. We then apply the greedy swapping algorithm described above to rearrange the frequency allocations. The resulting synchronization profiles are shown in Fig.~\ref{fig6} with blue curves and open circles. A substantial improvement in synchronization is clearly visible across all configurations, with the rearranged allocations achieving higher values of $R$ at lower coupling strengths compared to the random assignments. One can obtain similar improvements in synchronization for the frustrated case as well.

The results presented above demonstrate that a simple greedy algorithm for rearranging natural frequencies on a fixed network structure can substantially enhance synchronization in directed duplex networks. This improvement arises from aligning the frequency distribution with the underlying topology in a way that minimizes the MSAF. These findings naturally lead to a deeper question: what specific relationships between the dynamical and structural properties of the network are responsible for promoting synchronization? Addressing this question provides insight into the fundamental principles governing optimal network design.


\section{Structural–dynamical correlations in optimal duplex networks}

Having established that both network topology and frequency arrangement play crucial roles in achieving optimal synchronization, we now investigate the underlying relationships between structural and dynamical properties that promote synchronization in directed duplex networks. To address this question, we examine the system properties of synchrony-optimized duplex networks obtained through minimization of the multiplex synchronization alignment function (MSAF). We analyze three key correlations. In Sec.~\ref{freq-deg}, we study the correlations between oscillators' natural frequencies and their nodal in- and out-degrees in each layer. In~\ref{freq-freq}, we examine the correlations between the natural frequencies of neighboring oscillators within each layer. Finally, in~\ref{mirror-freq}, we investigate the correlations between the natural frequencies of mirror nodes across the two layers.

\subsection{Frequency–degree correlations}\label{freq-deg}

\begin{figure}[ht]
	\centering
	\includegraphics[width=\linewidth]{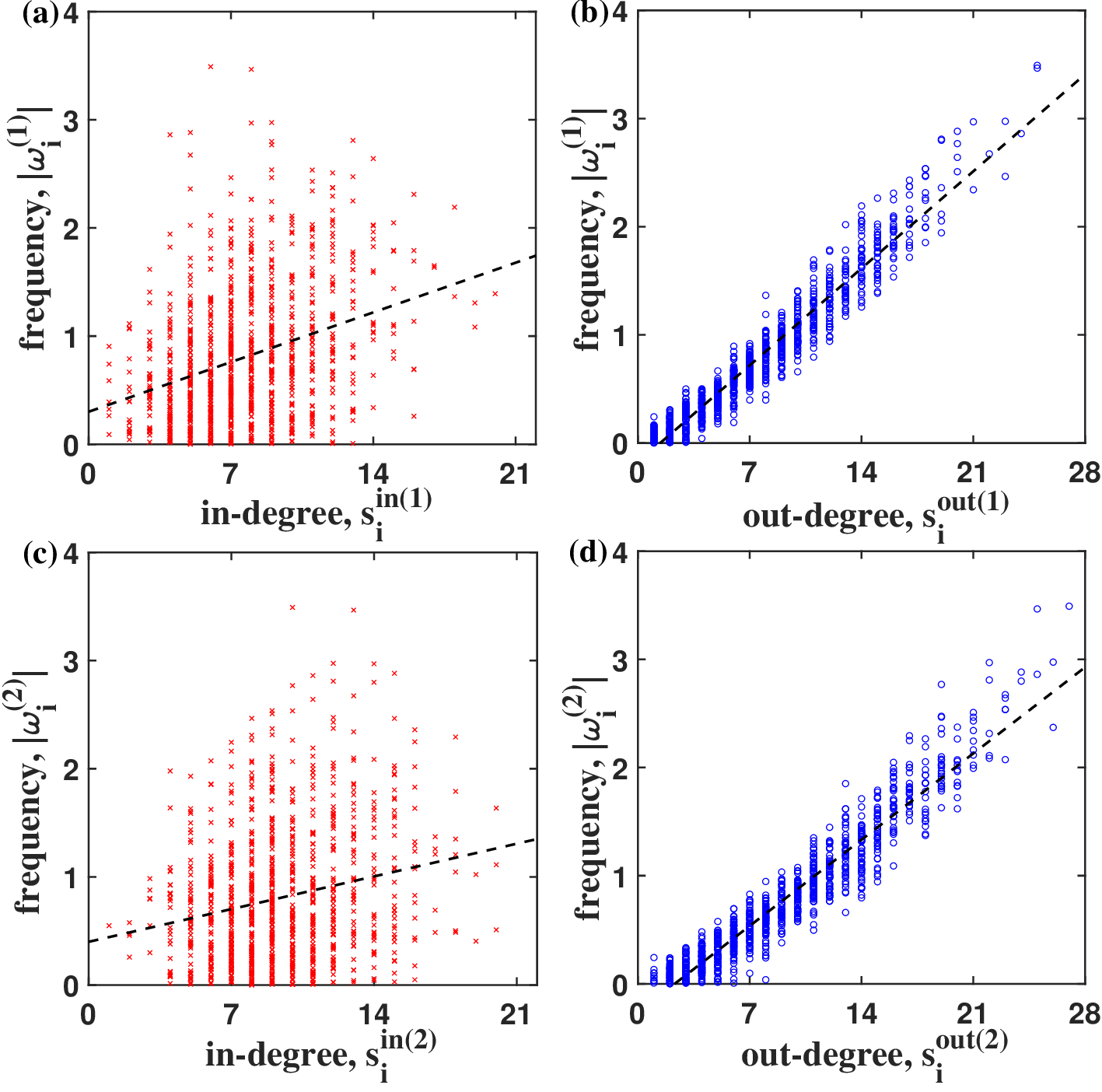}
	\caption{Frequency-degree correlations in an optimized ER-ER duplex network with normally distributed natural frequencies. (a) $\abs{\omega_{i}^{(1)}}$ vs.\ in-degree $s_{i}^{\mathrm{in}(1)}$ and (b) vs.\ out-degree $s_{i}^{\mathrm{out}(1)}$ for layer 1. (c) $\abs{\omega_{i}^{(2)}}$ vs.\ in-degree $s_{i}^{\mathrm{in}(2)}$ and (d) vs.\ out-degree $s_{i}^{\mathrm{out}(2)}$ for layer 2. Dashed black lines indicate least-squares linear fits. Pearson correlation coefficients: $C_{\pmb{s}^{\mathrm{in}(1)},\abs{\pmb{\omega}^{(1)}}}=0.3276$, $C_{\pmb{s}^{\mathrm{out}(1)},\abs{\pmb{\omega}^{(1)}}}=0.9666$, $C_{\pmb{s}^{\mathrm{in}(2)},\abs{\pmb{\omega}^{(2)}}}=0.2202$, $C_{\pmb{s}^{\mathrm{out}(2)},\abs{\pmb{\omega}^{(2)}}}=0.9551$.}
	\label{fig7}
\end{figure}

\begin{figure}[ht]
	\centering
	\includegraphics[width=\linewidth]{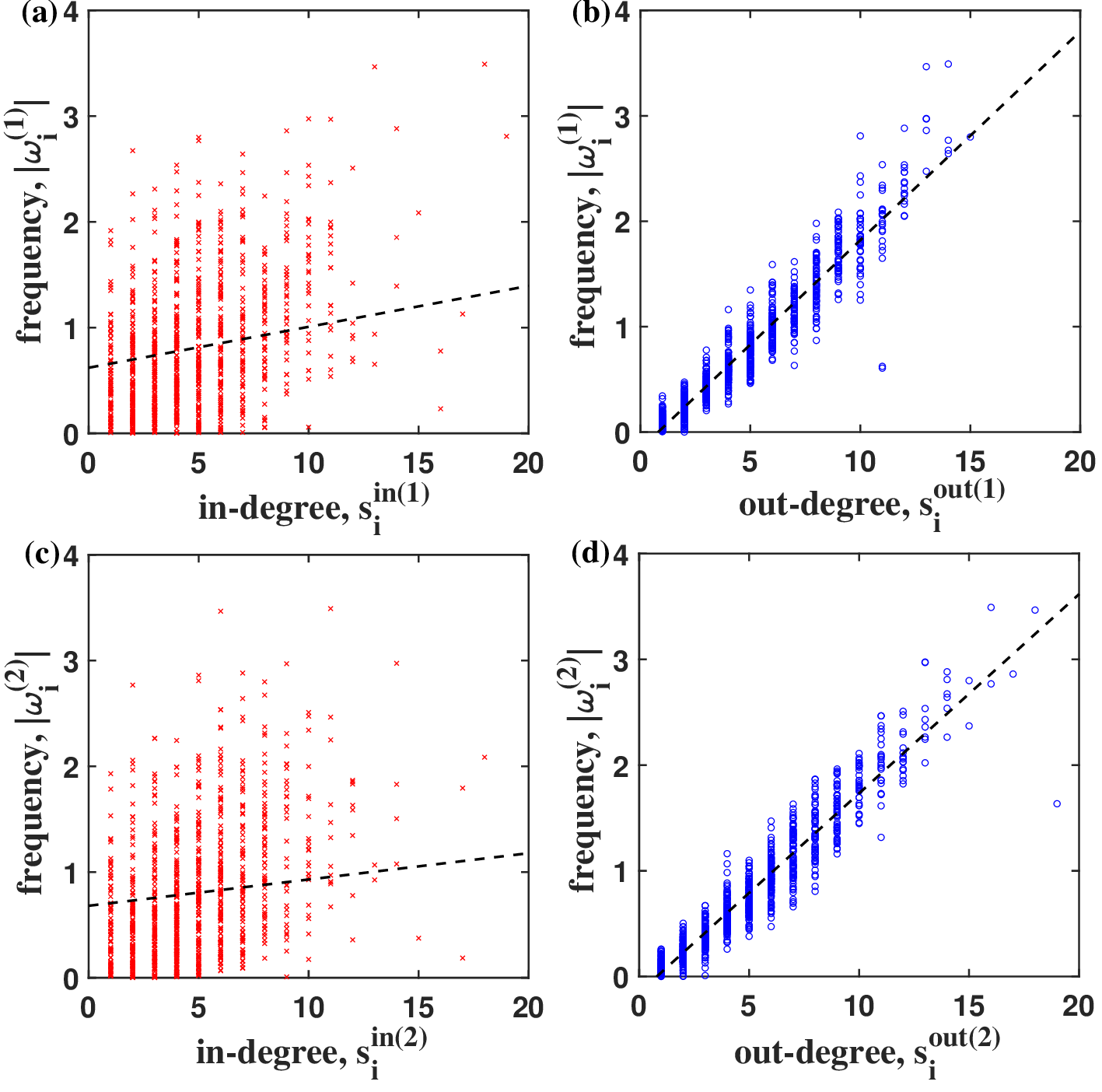}
	\caption{Frequency-degree correlations in an optimized SF-SF duplex network with normally distributed natural frequencies. (a) $\abs{\omega_{i}^{(1)}}$ vs.\ in-degree $s_{i}^{\mathrm{in}(1)}$ and (b) vs.\ out-degree $s_{i}^{\mathrm{out}(1)}$ for layer 1. (c) $\abs{\omega_{i}^{(2)}}$ vs.\ in-degree $s_{i}^{\mathrm{in}(2)}$ and (d) vs.\ out-degree $s_{i}^{\mathrm{out}(2)}$ for layer 2. Dashed black lines indicate least-squares linear fits. Pearson correlation coefficients: $C_{\pmb{s}^{\mathrm{in}(1)},\abs{\pmb{\omega}^{(1)}}}=0.2744$, $C_{\pmb{s}^{\mathrm{out}(1)},\abs{\pmb{\omega}^{(1)}}}=0.9532$, $C_{\pmb{s}^{\mathrm{in}(2)},\abs{\pmb{\omega}^{(2)}}}=0.2148$, $C_{\pmb{s}^{\mathrm{out}(2)},\abs{\pmb{\omega}^{(2)}}}=0.9544$.}
	\label{fig8}
\end{figure}

We begin our investigation by examining the correlations between the natural frequencies of oscillators and their structural connectivity in each layer. Specifically, we consider the relationship between the absolute value of the natural frequency $\abs{\omega_{i}^{(l)}}$ and the nodal in-degree $s_{i}^{\mathrm{in}(l)}$ and out-degree $s_{i}^{\mathrm{out}(l)}$ for $l=1,2$. The results presented here are obtained from synchrony-optimized networks generated through the rewiring algorithm described in Sec.~\ref{design-networks}, where a fixed set of pre-chosen normally distributed frequencies is assigned to the nodes. We note, however, that the observed trends are consistent across the other optimization scenarios considered in this study.

Figure~\ref{fig7} displays scatter plots for an optimized ER-ER duplex network. Panels (a) and (b) show $\abs{\omega_{i}^{(1)}}$ as a function of $s_{i}^{\mathrm{in}(1)}$ and $s_{i}^{\mathrm{out}(1)}$ for layer 1, respectively. Panels (c) and (d) present the corresponding results for layer 2, with $\abs{\omega_{i}^{(2)}}$ plotted against $s_{i}^{\mathrm{in}(2)}$ and $s_{i}^{\mathrm{out}(2)}$. In each panel, the dashed black line represents the least-squares linear fit. From these scatter plots, we observe a positive correlation between the natural frequencies and both in- and out-degrees in each layer. Notably, the relationship is considerably stronger for out-degrees than for in-degrees, suggesting that the optimization process preferentially aligns higher frequencies with nodes of higher out-degree. To quantify these observations, we compute the Pearson correlation coefficient between each pair of quantities,
\begin{equation}
	C_{\pmb{s},\abs{\pmb{\omega}}}=\frac{\sum_{i}(s_{i}-\innerproduct{\pmb{s}})(\abs{\omega_{i}}-\innerproduct{\abs{\pmb{\omega}}})}{\sqrt{\left[\sum_{i}(s_{i}-\innerproduct{\pmb{s}})^2\right]\left[\sum_{i}(\abs{\omega_{i}}-\innerproduct{\abs{\pmb{\omega}}})^2\right]}}. \label{eqn17}
\end{equation}
For the ER–ER network shown in Fig.~\ref{fig7}, the coefficients are $C_{\pmb{s}^{\mathrm{in}(1)},\abs{\pmb{\omega}^{(1)}}}=0.3276$, $C_{\pmb{s}^{\mathrm{out}(1)},\abs{\pmb{\omega}^{(1)}}}=0.9666$ for layer 1, and $C_{\pmb{s}^{\mathrm{in}(2)},\abs{\pmb{\omega}^{(2)}}}=0.2202$, $C_{\pmb{s}^{\mathrm{out}(2)},\abs{\pmb{\omega}^{(2)}}}=0.9551$ for layer 2. These values confirm the strong positive correlation with out-degree, while the correlation with in-degree is comparatively weak.

We next examine an optimized SF–SF duplex network to assess the generality of these findings. Figure~\ref{fig8} presents the corresponding scatter plots for the SF–SF configuration, with the same arrangement of panels as in Fig.~\ref{fig7}. Similar trends are evident: natural frequencies exhibit a positive relationship with both in- and out-degrees, with a markedly stronger correlation observed for out-degrees. The Pearson correlation coefficients for the SF–SF network are $C_{\pmb{s}^{\mathrm{in}(1)},\abs{\pmb{\omega}^{(1)}}}=0.2744$, $C_{\pmb{s}^{\mathrm{out}(1)},\abs{\pmb{\omega}^{(1)}}}=0.9532$ for layer 1, and $C_{\pmb{s}^{\mathrm{in}(2)},\abs{\pmb{\omega}^{(2)}}}=0.2148$, $C_{\pmb{s}^{\mathrm{out}(2)},\abs{\pmb{\omega}^{(2)}}}=0.9544$ for layer 2. These results are in close agreement with those obtained for the ER–ER network, indicating that the observed frequency–degree correlations are a robust feature of synchrony-optimized directed duplex networks, independent of the specific network topology.

\subsection{Frequency-frequency correlations}\label{freq-freq}

We next examine the relationship between natural frequencies of neighboring oscillators within each layer. Specifically, we investigate the correlation between an oscillator's natural frequency and the mean frequency of its neighbors, separately considering incoming and outgoing connections. To quantify these relationships, we define for layer~1 the mean in-frequency and mean out-frequency as
\begin{equation}
	{\innerproduct{\omega^{(1)}}}_{i}^{\mathrm{in}}=\frac{1}{s_{i}^{\mathrm{in}(1)}} \sum_{j=1}^{N} A_{ji}^{(1)}\omega_{j}^{(1)},~{\innerproduct{\omega^{(1)}}}_{i}^{\mathrm{out}}=\frac{1}{s_{i}^{\mathrm{out}(1)}} \sum_{j=1}^{N} A_{ji}^{(1)}\omega_{j}^{(1)}, \label{eqn18}
\end{equation}
and analogously for layer~2:
\begin{equation}
	{\innerproduct{\omega^{(2)}}}_{i}^{\mathrm{in}}=\frac{1}{s_{i}^{\mathrm{in}(2)}} \sum_{j=1}^{N} A_{ji}^{(2)}\omega_{j}^{(2)},~{\innerproduct{\omega^{(2)}}}_{i}^{\mathrm{out}}=\frac{1}{s_{i}^{\mathrm{out}(2)}} \sum_{j=1}^{N} A_{ji}^{(2)}\omega_{j}^{(2)}. \label{eqn19}
\end{equation}
Physically, ${\innerproduct{\omega^{(l)}}}_{i}^{\mathrm{in}}$ represents the average frequency of oscillators that influence node $i$ in layer $l$ (i.e., those with incoming links to $i$), while ${\innerproduct{\omega^{(l)}}}_{i}^{\mathrm{out}}$ denotes the average frequency of oscillators that are influenced by node $i$ in layer $l$ (i.e., those receiving outgoing links from $i$). These quantities thus capture the local frequency environment experienced by each oscillator through its directed connections.

\begin{figure}[ht]
	\centering
	\includegraphics[width=\linewidth]{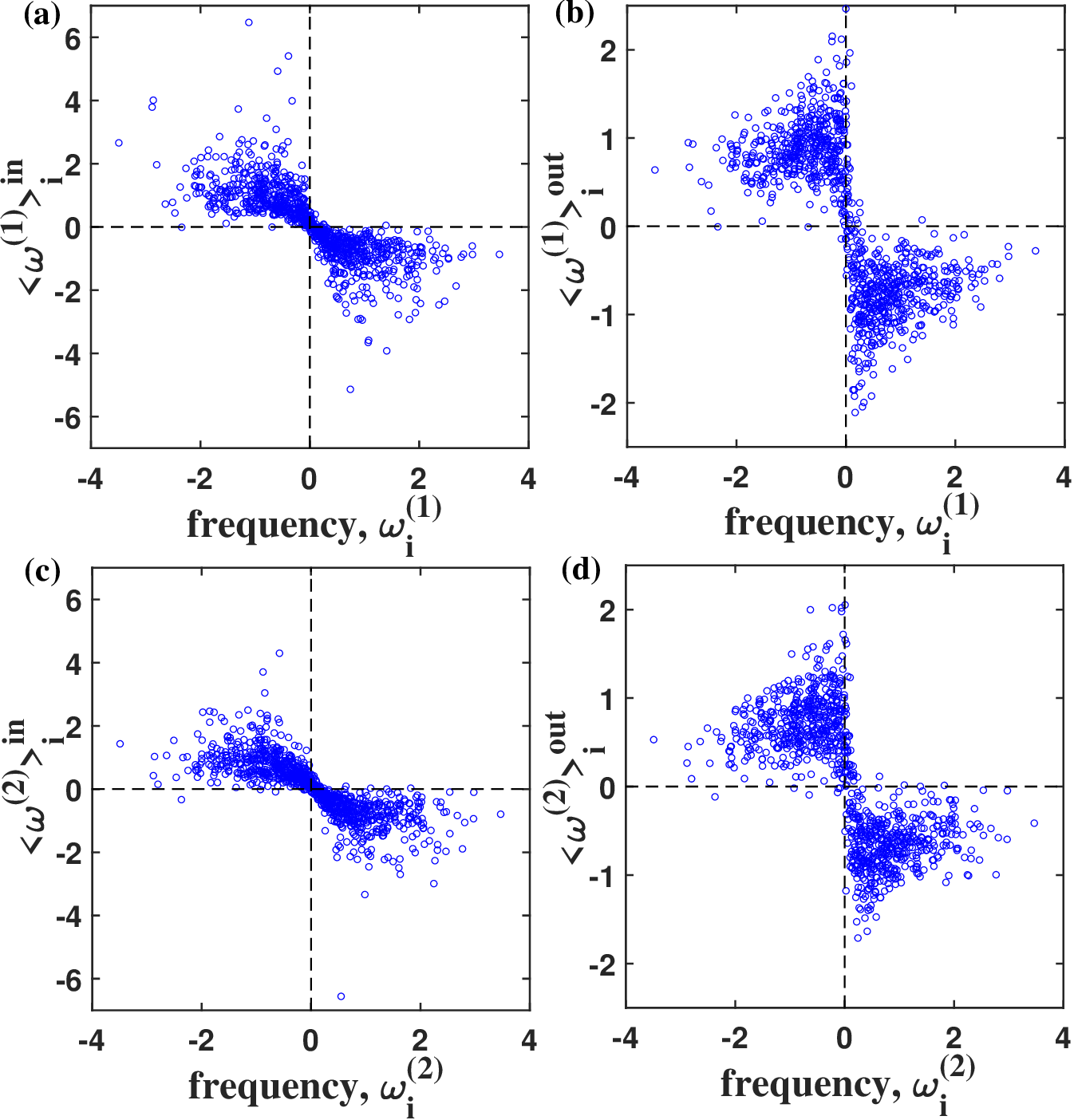}
	\caption{Frequency–frequency correlations in an optimized ER–ER duplex network with normally distributed natural frequencies. (a) Mean in-frequency $\innerproduct{\omega^{(1)}}_{i}^{\mathrm{in}}$ and (b) mean out-frequency $\innerproduct{\omega^{(1)}}_{i}^{\mathrm{out}}$ vs. natural frequency $\omega_{i}^{(1)}$ for layer 1. (c) Mean in-frequency $\innerproduct{\omega^{(2)}}_{i}^{\mathrm{in}}$ and (d) mean out-frequency $\innerproduct{\omega^{(2)}}_{i}^{\mathrm{out}}$ vs. natural frequency $\omega_{i}^{(2)}$ for layer 2.}
	\label{fig9}
\end{figure}

We analyze these correlations using synchrony-optimized networks generated through the rewiring algorithm described in Sec.~\ref{design-networks}, with normally distributed natural frequencies fixed in both layers. Figure~\ref{fig9} presents the results for an optimized ER–ER duplex network. Panels (a) and (b) show $\omega_{i}^{(1)}$ plotted against ${\innerproduct{\omega^{(1)}}}_{i}^{\mathrm{in}}$ and ${\innerproduct{\omega^{(1)}}}_{i}^{\mathrm{out}}$ for layer~1, respectively. Panels (c) and (d) display the corresponding quantities for layer~2. In all panels, a clear negative correlation is evident: oscillators with higher natural frequencies tend to have neighbors with lower mean frequencies, and vice versa. Notably, this negative relationship holds for both incoming and outgoing neighbor averages, indicating a consistent pattern of local frequency balancing regardless of the direction of influence.

\begin{figure}[ht]
	\centering
	\includegraphics[width=\linewidth]{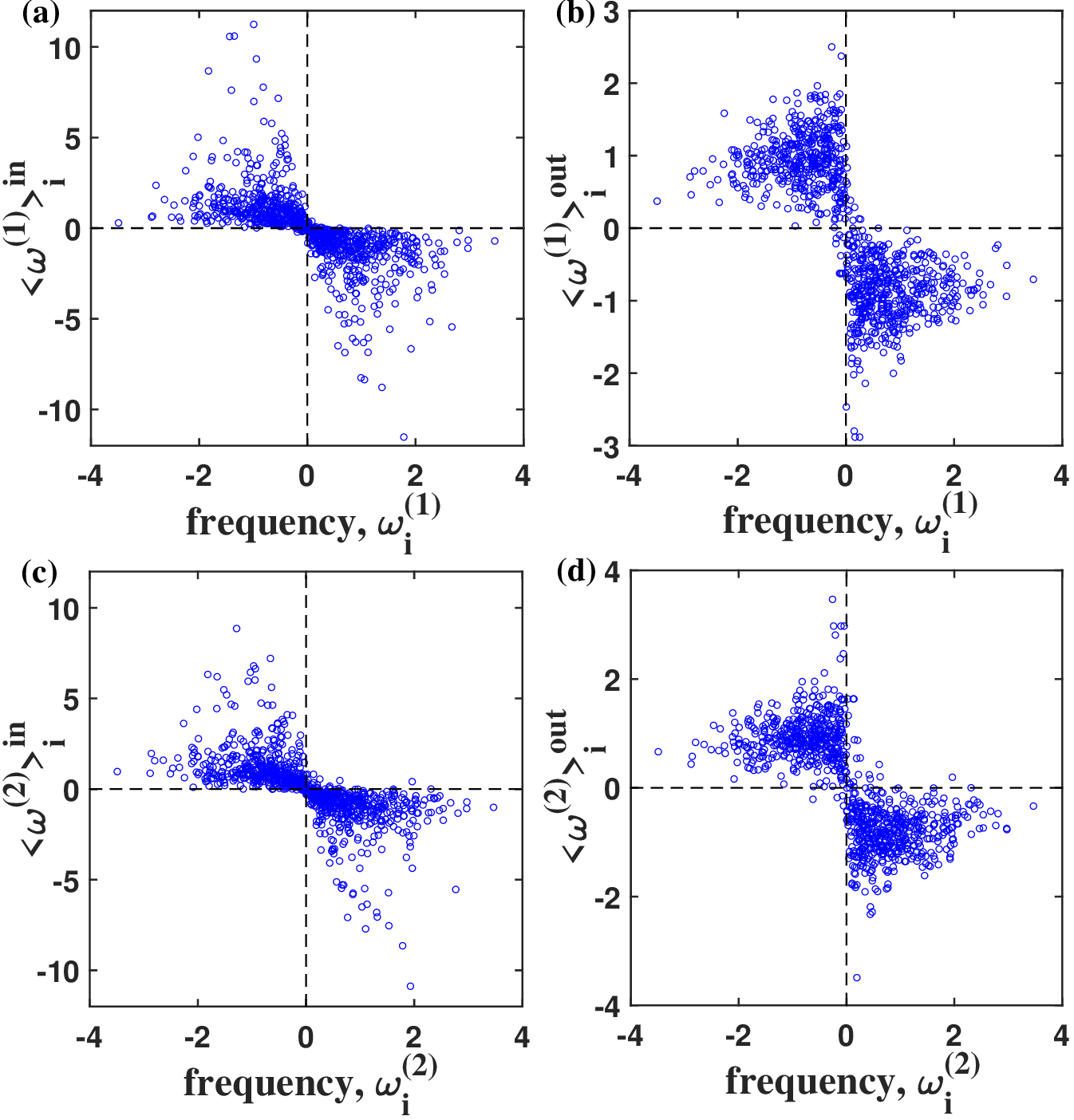}
	\caption{Frequency–frequency correlations in an optimized SF-SF duplex network with normally distributed natural frequencies. (a) Mean in-frequency $\innerproduct{\omega^{(1)}}_{i}^{\mathrm{in}}$ and (b) mean out-frequency $\innerproduct{\omega^{(1)}}_{i}^{\mathrm{out}}$ vs. natural frequency $\omega_{i}^{(1)}$ for layer 1. (c) Mean in-frequency $\innerproduct{\omega^{(2)}}_{i}^{\mathrm{in}}$ and (d) mean out-frequency $\innerproduct{\omega^{(2)}}_{i}^{\mathrm{out}}$ vs. natural frequency $\omega_{i}^{(2)}$ for layer 2.}
	\label{fig10}
\end{figure}

To assess the generality of this observation, we perform the same analysis on an optimized SF–SF duplex network, with results shown in Fig.~\ref{fig10}. The layout follows the same structure as Fig.~\ref{fig9}, with panels (a) and (b) for layer~1 and panels (c) and (d) for layer~2. Once again, a pronounced negative relationship is observed across all panels, consistent with the findings for the ER–ER network. This consistency indicates that the negative frequency–frequency correlation is a robust feature of synchrony-optimized directed duplex networks, independent of the underlying topological structure and similar for both in- and out-neighbor averages.

\subsection{Mirror nodes' frequency correlations}\label{mirror-freq}

\begin{figure}[ht]
	\centering
	\includegraphics[width=\linewidth]{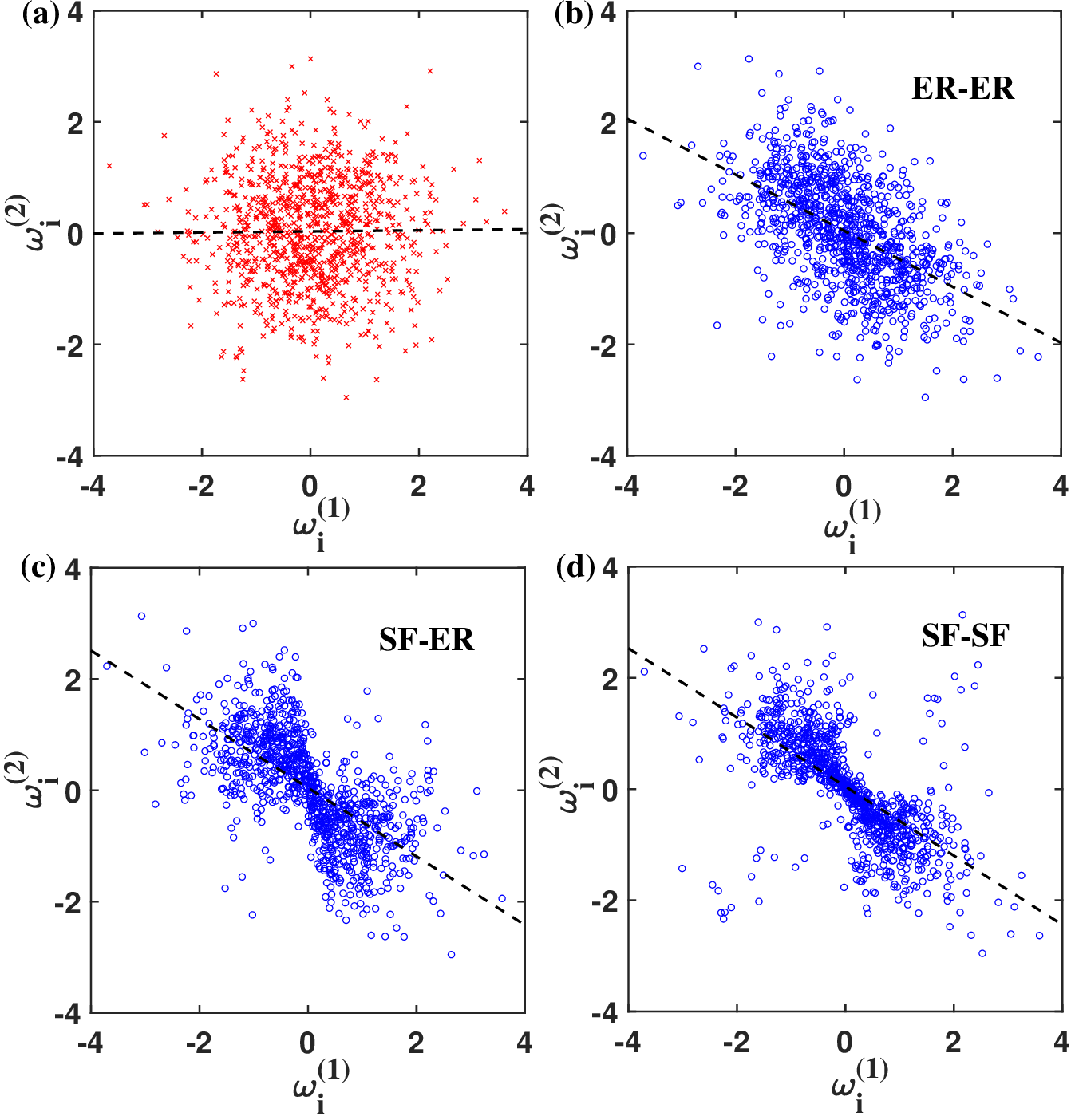}
	\caption{Mirror node frequency correlations in synchrony-optimized duplex networks. (a) Scatter plot of $\omega_{i}^{(1)}$ vs. $\omega_{i}^{(2)}$ for the initial normally distributed frequencies. (b) ER-ER, (c) SF-ER, and (d) SF-SF configurations after optimal rearrangement of oscillators in both layers. Dashed black lines indicate least-squares linear fits. Pearson correlation coefficients: (a) $0.0098$, (b) $-0.5035$, (c) $-0.6168$, (d) $-0.6214$.}		
	\label{fig11}
\end{figure}

Finally, we investigate the correlation between the natural frequencies of mirror nodes---pairs of oscillators representing the same entity across the two layers of the duplex network. The results presented in this section are obtained from optimally rearranged oscillators on fixed duplex networks, following the greedy algorithm described in Sec.~\ref{freq-arrange}. In this procedure, the natural frequencies in both layers are rearranged simultaneously to minimize the multiplex synchronization alignment function (MSAF) to a desired threshold, thereby enhancing global synchronization.

Figure~\ref{fig11} displays scatter plots of $\omega_{i}^{(1)}$ versus $\omega_{i}^{(2)}$ for various configurations. Panel (a) shows the reference case where normally distributed frequencies are allocated randomly to the oscillators in each layer without any rearrangement. As evident from the scatter plot, there is no discernible correlation between the mirror node frequencies in this initial configuration, indicating that the random allocation yields independent frequency assignments across the two layers. Panels (b), (c), and (d) present the results after optimal rearrangement for three duplex configurations: ER-ER (b), SF-ER (c), and SF-SF (d). In each panel, the dashed black line represents the least-squares linear fit to the scatter plot. Strikingly, after rearrangement, a clear negative correlation emerges between the natural frequencies of mirror nodes in all three configurations. This observation suggests that the optimization process favors arrangements where a high-frequency oscillator in one layer is paired with a low-frequency mirror node in the other layer, and vice versa. Such anti-correlation across layers likely contributes to the enhanced global synchronization by promoting balanced frequency distributions in the interlayer coupling terms. To quantify these observations, we compute the Pearson correlation coefficient between $\omega_{i}^{(1)}$ and $\omega_{i}^{(2)}$ for each panel, defined as
\begin{equation}
	C_{\pmb{\omega}^{(1)},\pmb{\omega}^{(2)}}=\frac{\sum_{i} (\omega_{i}^{(1)}-\innerproduct{\pmb{\omega}^{(1)}})(\omega_{i}^{(2)}-\innerproduct{\pmb{\omega}^{(2)}})}{\sqrt{\left[\sum_{i}(\omega_{i}^{(1)}-\innerproduct{\pmb{\omega}^{(1)}})^2\right]\left[\sum_{i}(\omega_{i}^{(2)}-\innerproduct{\pmb{\omega}^{(2)}})^2\right]}}. \label{eqn20}
\end{equation}
For the four panels shown in Fig.~\ref{fig11}, the computed correlation coefficients are as follows: panel (a) yields $C_{\pmb{\omega}^{(1)},\pmb{\omega}^{(2)}}=0.0098$, indicating essentially no such correlation in the random initial allocation. After optimal rearrangement, panel (b) (ER-ER) gives $C_{\pmb{\omega}^{(1)},\pmb{\omega}^{(2)}}=-0.5035$, panel (c) (SF-ER) gives $C_{\pmb{\omega}^{(1)},\pmb{\omega}^{(2)}}=-0.6168$, and panel (d) (SF-SF) gives $C_{\pmb{\omega}^{(1)},\pmb{\omega}^{(2)}}=-0.6214$. These negative values confirm the presence of a moderate to strong anti-correlation between mirror node frequencies in synchrony-optimized duplex networks.

Together with the findings from the preceding subsections, these results reveal a consistent set of structural–dynamical signatures that characterize optimal synchronization in directed duplex networks: high-frequency oscillators tend to possess high out-degree, are surrounded by low-frequency neighbors, and are paired with low-frequency mirror nodes in the other layer. This multi-faceted alignment between frequencies and network structure collectively facilitates the emergence of coherent collective dynamics. With this understanding, we now summarize our main findings and discuss their implications in the concluding section that follows.


\section{Conclusions}

In this work, we have developed a comprehensive analytical and numerical framework for understanding and enhancing synchronization in directed multiplex networks, with a focus on duplex configurations consisting of two interconnected layers. By extending the Sakaguchi-Kuramoto model to incorporate directed intra-layer connectivity and layer-specific phase-lag parameters, we have derived the multiplex synchronization alignment function (MSAF) as a key metric for quantifying the interplay between network topology, frequency distribution, and synchronization quality. Our analytical framework yielded two distinct frequency sets for enhancing synchronization. The perfect frequency sets achieve perfect synchronization at a targeted coupling strength $K_{\mathrm{p}}$ and depend linearly on the out-degree distribution and frustration parameter of each layer. The optimal frequency sets provide enhanced synchronization across a broad range of coupling strengths and explicitly depend on $K$ in the presence of frustration. Numerical validation across three directed duplex configurations---SF-SF, SF-ER, and ER-ER---confirmed the effectiveness of both frequency sets. In the non-frustrated regime, the optimal frequencies enabled synchronization at significantly lower coupling strengths compared to standard uniform, normal, and Lorentzian distributions. In the frustrated regime, both the optimal and perfect frequency sets successfully enhanced synchronization, with the perfect sets achieving perfect coherence across both layers at the prescribed $K_{\mathrm{p}}$.

We further addressed the inverse problem of network design, demonstrating that simultaneous rewiring of both layers guided by MSAF minimization leads to substantial improvements in global synchronization. Additionally, a greedy algorithm for optimal frequency arrangement on fixed networks showed that rearranging frequencies to minimize the MSAF significantly enhances synchronization compared to random allocation. Our investigation of structural-dynamical correlations revealed three characteristic signatures of synchrony-optimized networks: a strong positive correlation between natural frequencies and nodal out-degree, a consistent negative correlation between frequencies of neighboring oscillators, and a negative correlation between mirror node frequencies across the two layers. These findings paint a coherent picture where high-frequency oscillators possess high out-degree, are surrounded by low-frequency neighbors, and are paired with low-frequency mirror nodes. Building on these results, a natural direction for future research is to extend the proposed optimization framework to higher-order interactions, such as directed simplicial complexes~\cite{bjorner1999complexes} or directed hypergraphs~\cite{gallo1993directed}, where synchronization dynamics are governed not only by pairwise but also by group-wise couplings. Such an extension would enable a more faithful representation of complex systems with multi-body interactions and could further enrich the understanding of synchronization phenomena in realistic settings. From an application perspective, the proposed framework offers valuable insights for the design and control of power grids~\cite{motter2013spontaneous,menck2014dead}, where generators can be modeled as phase oscillators and appropriately designed frequency configurations may help prevent cascading failures by ensuring stable synchronization under varying load conditions.


\section*{Data Availability}

The data that support the findings of this study are available within the article.


\section*{Author Declarations}

\subsection*{Conflict of Interest}

The authors have no conflicts to disclose.


%


\section*{References}
\begin{thebibliography}{66}%
	\makeatletter
	\providecommand \@ifxundefined [1]{%
		\@ifx{#1\undefined}
	}%
	\providecommand \@ifnum [1]{%
		\ifnum #1\expandafter \@firstoftwo
		\else \expandafter \@secondoftwo
		\fi
	}%
	\providecommand \@ifx [1]{%
		\ifx #1\expandafter \@firstoftwo
		\else \expandafter \@secondoftwo
		\fi
	}%
	\providecommand \natexlab [1]{#1}%
	\providecommand \enquote  [1]{``#1''}%
	\providecommand \bibnamefont  [1]{#1}%
	\providecommand \bibfnamefont [1]{#1}%
	\providecommand \citenamefont [1]{#1}%
	\providecommand \href@noop [0]{\@secondoftwo}%
	\providecommand \href [0]{\begingroup \@sanitize@url \@href}%
	\providecommand \@href[1]{\@@startlink{#1}\@@href}%
	\providecommand \@@href[1]{\endgroup#1\@@endlink}%
	\providecommand \@sanitize@url [0]{\catcode `\\12\catcode `\$12\catcode
		`\&12\catcode `\#12\catcode `\^12\catcode `\_12\catcode `\%12\relax}%
	\providecommand \@@startlink[1]{}%
	\providecommand \@@endlink[0]{}%
	\providecommand \url  [0]{\begingroup\@sanitize@url \@url }%
	\providecommand \@url [1]{\endgroup\@href {#1}{\urlprefix }}%
	\providecommand \urlprefix  [0]{URL }%
	\providecommand \Eprint [0]{\href }%
	\providecommand \doibase [0]{http://dx.doi.org/}%
	\providecommand \selectlanguage [0]{\@gobble}%
	\providecommand \bibinfo  [0]{\@secondoftwo}%
	\providecommand \bibfield  [0]{\@secondoftwo}%
	\providecommand \translation [1]{[#1]}%
	\providecommand \BibitemOpen [0]{}%
	\providecommand \bibitemStop [0]{}%
	\providecommand \bibitemNoStop [0]{.\EOS\space}%
	\providecommand \EOS [0]{\spacefactor3000\relax}%
	\providecommand \BibitemShut  [1]{\csname bibitem#1\endcsname}%
	\let\auto@bib@innerbib\@empty
	\bibitem [{\citenamefont {Boccaletti}\ \emph {et~al.}(2014)\citenamefont
		{Boccaletti}, \citenamefont {Bianconi}, \citenamefont {Criado}, \citenamefont
		{Del~Genio}, \citenamefont {G{\'o}mez-Gardenes}, \citenamefont {Romance},
		\citenamefont {Sendina-Nadal}, \citenamefont {Wang},\ and\ \citenamefont
		{Zanin}}]{boccaletti2014structure}%
	\BibitemOpen
	\bibfield  {author} {\bibinfo {author} {\bibfnamefont {S.}~\bibnamefont
			{Boccaletti}}, \bibinfo {author} {\bibfnamefont {G.}~\bibnamefont
			{Bianconi}}, \bibinfo {author} {\bibfnamefont {R.}~\bibnamefont {Criado}},
		\bibinfo {author} {\bibfnamefont {C.~I.}\ \bibnamefont {Del~Genio}}, \bibinfo
		{author} {\bibfnamefont {J.}~\bibnamefont {G{\'o}mez-Gardenes}}, \bibinfo
		{author} {\bibfnamefont {M.}~\bibnamefont {Romance}}, \bibinfo {author}
		{\bibfnamefont {I.}~\bibnamefont {Sendina-Nadal}}, \bibinfo {author}
		{\bibfnamefont {Z.}~\bibnamefont {Wang}}, \ and\ \bibinfo {author}
		{\bibfnamefont {M.}~\bibnamefont {Zanin}},\ }\bibfield  {title} {\enquote
		{\bibinfo {title} {The structure and dynamics of multilayer networks},}\
	}\href@noop {} {\bibfield  {journal} {\bibinfo  {journal} {Physics Reports}\
		}\textbf {\bibinfo {volume} {544}},\ \bibinfo {pages} {1--122} (\bibinfo
		{year} {2014})}\BibitemShut {NoStop}%
	\bibitem [{\citenamefont {Bianconi}(2018)}]{bianconi2018multilayer}%
	\BibitemOpen
	\bibfield  {author} {\bibinfo {author} {\bibfnamefont {G.}~\bibnamefont
			{Bianconi}},\ }\href@noop {} {\emph {\bibinfo {title} {Multilayer Networks:
				Structure and Function}}}\ (\bibinfo  {publisher} {Oxford University Press},\
	\bibinfo {year} {2018})\BibitemShut {NoStop}%
	\bibitem [{\citenamefont {Albert}\ and\ \citenamefont
		{Barab{\'a}si}(2002)}]{albert2002statistical}%
	\BibitemOpen
	\bibfield  {author} {\bibinfo {author} {\bibfnamefont {R.}~\bibnamefont
			{Albert}}\ and\ \bibinfo {author} {\bibfnamefont {A.-L.}\ \bibnamefont
			{Barab{\'a}si}},\ }\bibfield  {title} {\enquote {\bibinfo {title}
			{Statistical mechanics of complex networks},}\ }\href@noop {} {\bibfield
		{journal} {\bibinfo  {journal} {Reviews of Modern Physics}\ }\textbf
		{\bibinfo {volume} {74}},\ \bibinfo {pages} {47} (\bibinfo {year}
		{2002})}\BibitemShut {NoStop}%
	\bibitem [{\citenamefont {Newman}(2003)}]{newman2003structure}%
	\BibitemOpen
	\bibfield  {author} {\bibinfo {author} {\bibfnamefont {M.~E.~J.}\
			\bibnamefont {Newman}},\ }\bibfield  {title} {\enquote {\bibinfo {title} {The
				structure and function of complex networks},}\ }\href@noop {} {\bibfield
		{journal} {\bibinfo  {journal} {SIAM Review}\ }\textbf {\bibinfo {volume}
			{45}},\ \bibinfo {pages} {167--256} (\bibinfo {year} {2003})}\BibitemShut
	{NoStop}%
	\bibitem [{\citenamefont {Boccaletti}\ \emph {et~al.}(2006)\citenamefont
		{Boccaletti}, \citenamefont {Latora}, \citenamefont {Moreno}, \citenamefont
		{Chavez},\ and\ \citenamefont {Hwang}}]{boccaletti2006complex}%
	\BibitemOpen
	\bibfield  {author} {\bibinfo {author} {\bibfnamefont {S.}~\bibnamefont
			{Boccaletti}}, \bibinfo {author} {\bibfnamefont {V.}~\bibnamefont {Latora}},
		\bibinfo {author} {\bibfnamefont {Y.}~\bibnamefont {Moreno}}, \bibinfo
		{author} {\bibfnamefont {M.}~\bibnamefont {Chavez}}, \ and\ \bibinfo {author}
		{\bibfnamefont {D.-U.}\ \bibnamefont {Hwang}},\ }\bibfield  {title} {\enquote
		{\bibinfo {title} {Complex networks: Structure and dynamics},}\ }\href@noop
	{} {\bibfield  {journal} {\bibinfo  {journal} {Physics Reports}\ }\textbf
		{\bibinfo {volume} {424}},\ \bibinfo {pages} {175--308} (\bibinfo {year}
		{2006})}\BibitemShut {NoStop}%
	\bibitem [{\citenamefont {Aleta}\ and\ \citenamefont
		{Moreno}(2019)}]{aleta2019multilayer}%
	\BibitemOpen
	\bibfield  {author} {\bibinfo {author} {\bibfnamefont {A.}~\bibnamefont
			{Aleta}}\ and\ \bibinfo {author} {\bibfnamefont {Y.}~\bibnamefont {Moreno}},\
	}\bibfield  {title} {\enquote {\bibinfo {title} {Multilayer networks in a
				nutshell},}\ }\href@noop {} {\bibfield  {journal} {\bibinfo  {journal}
			{Annual Review of Condensed Matter Physics}\ }\textbf {\bibinfo {volume}
			{10}},\ \bibinfo {pages} {45--62} (\bibinfo {year} {2019})}\BibitemShut
	{NoStop}%
	\bibitem [{\citenamefont {Myers}, \citenamefont {Zhu},\ and\ \citenamefont
		{Leskovec}(2012)}]{myers2012information}%
	\BibitemOpen
	\bibfield  {author} {\bibinfo {author} {\bibfnamefont {S.~A.}\ \bibnamefont
			{Myers}}, \bibinfo {author} {\bibfnamefont {C.}~\bibnamefont {Zhu}}, \ and\
		\bibinfo {author} {\bibfnamefont {J.}~\bibnamefont {Leskovec}},\ }\bibfield
	{title} {\enquote {\bibinfo {title} {Information diffusion and external
				influence in networks},}\ }in\ \href@noop {} {\emph {\bibinfo {booktitle}
			{Proceedings of the 18th ACM SIGKDD International Conference on Knowledge
				Discovery and Data Mining}}}\ (\bibinfo {organization} {New York: ACM},\
	\bibinfo {year} {2012})\ pp.\ \bibinfo {pages} {33--41}\BibitemShut {NoStop}%
	\bibitem [{\citenamefont {Vicsek}\ \emph {et~al.}(1995)\citenamefont {Vicsek},
		\citenamefont {Czir{\'o}k}, \citenamefont {Ben-Jacob}, \citenamefont
		{Cohen},\ and\ \citenamefont {Shochet}}]{vicsek1995novel}%
	\BibitemOpen
	\bibfield  {author} {\bibinfo {author} {\bibfnamefont {T.}~\bibnamefont
			{Vicsek}}, \bibinfo {author} {\bibfnamefont {A.}~\bibnamefont {Czir{\'o}k}},
		\bibinfo {author} {\bibfnamefont {E.}~\bibnamefont {Ben-Jacob}}, \bibinfo
		{author} {\bibfnamefont {I.}~\bibnamefont {Cohen}}, \ and\ \bibinfo {author}
		{\bibfnamefont {O.}~\bibnamefont {Shochet}},\ }\bibfield  {title} {\enquote
		{\bibinfo {title} {Novel type of phase transition in a system of self-driven
				particles},}\ }\href@noop {} {\bibfield  {journal} {\bibinfo  {journal}
			{Physical Review Letters}\ }\textbf {\bibinfo {volume} {75}},\ \bibinfo
		{pages} {1226} (\bibinfo {year} {1995})}\BibitemShut {NoStop}%
	\bibitem [{\citenamefont {Prorok}, \citenamefont {Hsieh},\ and\ \citenamefont
		{Kumar}(2017)}]{prorok2017impact}%
	\BibitemOpen
	\bibfield  {author} {\bibinfo {author} {\bibfnamefont {A.}~\bibnamefont
			{Prorok}}, \bibinfo {author} {\bibfnamefont {M.~A.}\ \bibnamefont {Hsieh}}, \
		and\ \bibinfo {author} {\bibfnamefont {V.}~\bibnamefont {Kumar}},\ }\bibfield
	{title} {\enquote {\bibinfo {title} {The impact of diversity on optimal
				control policies for heterogeneous robot swarms},}\ }\href@noop {} {\bibfield
		{journal} {\bibinfo  {journal} {IEEE Transactions on Robotics}\ }\textbf
		{\bibinfo {volume} {33}},\ \bibinfo {pages} {346--358} (\bibinfo {year}
		{2017})}\BibitemShut {NoStop}%
	\bibitem [{\citenamefont {Vaiana}\ and\ \citenamefont
		{Muldoon}(2020)}]{vaiana2020multilayer}%
	\BibitemOpen
	\bibfield  {author} {\bibinfo {author} {\bibfnamefont {M.}~\bibnamefont
			{Vaiana}}\ and\ \bibinfo {author} {\bibfnamefont {S.~F.}\ \bibnamefont
			{Muldoon}},\ }\bibfield  {title} {\enquote {\bibinfo {title} {Multilayer
				brain networks},}\ }\href@noop {} {\bibfield  {journal} {\bibinfo  {journal}
			{Journal of Nonlinear Science}\ }\textbf {\bibinfo {volume} {30}},\ \bibinfo
		{pages} {2147--2169} (\bibinfo {year} {2020})}\BibitemShut {NoStop}%
	\bibitem [{\citenamefont {Lim}\ \emph {et~al.}(2019)\citenamefont {Lim},
		\citenamefont {Radicchi}, \citenamefont {van~den Heuvel},\ and\ \citenamefont
		{Sporns}}]{lim2019discordant}%
	\BibitemOpen
	\bibfield  {author} {\bibinfo {author} {\bibfnamefont {S.}~\bibnamefont
			{Lim}}, \bibinfo {author} {\bibfnamefont {F.}~\bibnamefont {Radicchi}},
		\bibinfo {author} {\bibfnamefont {M.~P.}\ \bibnamefont {van~den Heuvel}}, \
		and\ \bibinfo {author} {\bibfnamefont {O.}~\bibnamefont {Sporns}},\
	}\bibfield  {title} {\enquote {\bibinfo {title} {Discordant attributes of
				structural and functional brain connectivity in a two-layer multiplex
				network},}\ }\href@noop {} {\bibfield  {journal} {\bibinfo  {journal}
			{Scientific Reports}\ }\textbf {\bibinfo {volume} {9}},\ \bibinfo {pages}
		{2885} (\bibinfo {year} {2019})}\BibitemShut {NoStop}%
	\bibitem [{\citenamefont {Battiston}\ \emph {et~al.}(2017)\citenamefont
		{Battiston}, \citenamefont {Nicosia}, \citenamefont {Chavez},\ and\
		\citenamefont {Latora}}]{battiston2017multilayer}%
	\BibitemOpen
	\bibfield  {author} {\bibinfo {author} {\bibfnamefont {F.}~\bibnamefont
			{Battiston}}, \bibinfo {author} {\bibfnamefont {V.}~\bibnamefont {Nicosia}},
		\bibinfo {author} {\bibfnamefont {M.}~\bibnamefont {Chavez}}, \ and\ \bibinfo
		{author} {\bibfnamefont {V.}~\bibnamefont {Latora}},\ }\bibfield  {title}
	{\enquote {\bibinfo {title} {Multilayer motif analysis of brain networks},}\
	}\href@noop {} {\bibfield  {journal} {\bibinfo  {journal} {Chaos: An
				Interdisciplinary Journal of Nonlinear Science}\ }\textbf {\bibinfo {volume}
			{27}},\ \bibinfo {pages} {047404} (\bibinfo {year} {2017})}\BibitemShut
	{NoStop}%
	\bibitem [{\citenamefont {Wei}\ \emph {et~al.}(2016)\citenamefont {Wei},
		\citenamefont {Chen}, \citenamefont {Wu}, \citenamefont {Ning},\ and\
		\citenamefont {Lu}}]{wei2016cooperative}%
	\BibitemOpen
	\bibfield  {author} {\bibinfo {author} {\bibfnamefont {X.}~\bibnamefont
			{Wei}}, \bibinfo {author} {\bibfnamefont {S.}~\bibnamefont {Chen}}, \bibinfo
		{author} {\bibfnamefont {X.}~\bibnamefont {Wu}}, \bibinfo {author}
		{\bibfnamefont {D.}~\bibnamefont {Ning}}, \ and\ \bibinfo {author}
		{\bibfnamefont {J.-a.}\ \bibnamefont {Lu}},\ }\bibfield  {title} {\enquote
		{\bibinfo {title} {Cooperative spreading processes in multiplex networks},}\
	}\href@noop {} {\bibfield  {journal} {\bibinfo  {journal} {Chaos: An
				Interdisciplinary Journal of Nonlinear Science}\ }\textbf {\bibinfo {volume}
			{26}},\ \bibinfo {pages} {065311} (\bibinfo {year} {2016})}\BibitemShut
	{NoStop}%
	\bibitem [{\citenamefont {Brummitt}, \citenamefont {D’Souza},\ and\
		\citenamefont {Leicht}(2012)}]{brummitt2012suppressing}%
	\BibitemOpen
	\bibfield  {author} {\bibinfo {author} {\bibfnamefont {C.~D.}\ \bibnamefont
			{Brummitt}}, \bibinfo {author} {\bibfnamefont {R.~M.}\ \bibnamefont
			{D’Souza}}, \ and\ \bibinfo {author} {\bibfnamefont {E.~A.}\ \bibnamefont
			{Leicht}},\ }\bibfield  {title} {\enquote {\bibinfo {title} {Suppressing
				cascades of load in interdependent networks},}\ }\href@noop {} {\bibfield
		{journal} {\bibinfo  {journal} {Proceedings of the National Academy of
				Sciences}\ }\textbf {\bibinfo {volume} {109}},\ \bibinfo {pages} {E680--E689}
		(\bibinfo {year} {2012})}\BibitemShut {NoStop}%
	\bibitem [{\citenamefont {G{\'o}mez-Gardenes}\ \emph
		{et~al.}(2012)\citenamefont {G{\'o}mez-Gardenes}, \citenamefont {Reinares},
		\citenamefont {Arenas},\ and\ \citenamefont
		{Flor{\'\i}a}}]{gomez2012evolution}%
	\BibitemOpen
	\bibfield  {author} {\bibinfo {author} {\bibfnamefont {J.}~\bibnamefont
			{G{\'o}mez-Gardenes}}, \bibinfo {author} {\bibfnamefont {I.}~\bibnamefont
			{Reinares}}, \bibinfo {author} {\bibfnamefont {A.}~\bibnamefont {Arenas}}, \
		and\ \bibinfo {author} {\bibfnamefont {L.~M.}\ \bibnamefont {Flor{\'\i}a}},\
	}\bibfield  {title} {\enquote {\bibinfo {title} {Evolution of cooperation in
				multiplex networks},}\ }\href@noop {} {\bibfield  {journal} {\bibinfo
			{journal} {Scientific Reports}\ }\textbf {\bibinfo {volume} {2}},\ \bibinfo
		{pages} {620} (\bibinfo {year} {2012})}\BibitemShut {NoStop}%
	\bibitem [{\citenamefont {Perc}\ \emph {et~al.}(2013)\citenamefont {Perc},
		\citenamefont {G{\'o}mez-Gardenes}, \citenamefont {Szolnoki}, \citenamefont
		{Flor{\'\i}a},\ and\ \citenamefont {Moreno}}]{perc2013evolutionary}%
	\BibitemOpen
	\bibfield  {author} {\bibinfo {author} {\bibfnamefont {M.}~\bibnamefont
			{Perc}}, \bibinfo {author} {\bibfnamefont {J.}~\bibnamefont
			{G{\'o}mez-Gardenes}}, \bibinfo {author} {\bibfnamefont {A.}~\bibnamefont
			{Szolnoki}}, \bibinfo {author} {\bibfnamefont {L.~M.}\ \bibnamefont
			{Flor{\'\i}a}}, \ and\ \bibinfo {author} {\bibfnamefont {Y.}~\bibnamefont
			{Moreno}},\ }\bibfield  {title} {\enquote {\bibinfo {title} {Evolutionary
				dynamics of group interactions on structured populations: a review},}\
	}\href@noop {} {\bibfield  {journal} {\bibinfo  {journal} {Journal of the
				Royal Society Interface}\ }\textbf {\bibinfo {volume} {10}},\ \bibinfo
		{pages} {20120997} (\bibinfo {year} {2013})}\BibitemShut {NoStop}%
	\bibitem [{\citenamefont {Tadi{\'c}}(2001)}]{tadic2001dynamics}%
	\BibitemOpen
	\bibfield  {author} {\bibinfo {author} {\bibfnamefont {B.}~\bibnamefont
			{Tadi{\'c}}},\ }\bibfield  {title} {\enquote {\bibinfo {title} {Dynamics of
				directed graphs: the world-wide web},}\ }\href@noop {} {\bibfield  {journal}
		{\bibinfo  {journal} {Physica A: Statistical Mechanics and its Applications}\
		}\textbf {\bibinfo {volume} {293}},\ \bibinfo {pages} {273--284} (\bibinfo
		{year} {2001})}\BibitemShut {NoStop}%
	\bibitem [{\citenamefont {MacKay}, \citenamefont {Johnson},\ and\ \citenamefont
		{Sansom}(2020)}]{mackay2020directed}%
	\BibitemOpen
	\bibfield  {author} {\bibinfo {author} {\bibfnamefont {R.~S.}\ \bibnamefont
			{MacKay}}, \bibinfo {author} {\bibfnamefont {S.}~\bibnamefont {Johnson}}, \
		and\ \bibinfo {author} {\bibfnamefont {B.}~\bibnamefont {Sansom}},\
	}\bibfield  {title} {\enquote {\bibinfo {title} {How directed is a directed
				network?}}\ }\href@noop {} {\bibfield  {journal} {\bibinfo  {journal} {Royal
				Society Open Science}\ }\textbf {\bibinfo {volume} {7}},\ \bibinfo {pages}
		{201138} (\bibinfo {year} {2020})}\BibitemShut {NoStop}%
	\bibitem [{\citenamefont {Chen}\ \emph {et~al.}(2013)\citenamefont {Chen},
		\citenamefont {Gao}, \citenamefont {L{\"u}},\ and\ \citenamefont
		{Zhou}}]{chen2013identifying}%
	\BibitemOpen
	\bibfield  {author} {\bibinfo {author} {\bibfnamefont {D.-B.}\ \bibnamefont
			{Chen}}, \bibinfo {author} {\bibfnamefont {H.}~\bibnamefont {Gao}}, \bibinfo
		{author} {\bibfnamefont {L.}~\bibnamefont {L{\"u}}}, \ and\ \bibinfo {author}
		{\bibfnamefont {T.}~\bibnamefont {Zhou}},\ }\bibfield  {title} {\enquote
		{\bibinfo {title} {Identifying influential nodes in large-scale directed
				networks: the role of clustering},}\ }\href@noop {} {\bibfield  {journal}
		{\bibinfo  {journal} {PLoS One}\ }\textbf {\bibinfo {volume} {8}},\ \bibinfo
		{pages} {e77455} (\bibinfo {year} {2013})}\BibitemShut {NoStop}%
	\bibitem [{\citenamefont {Segal}\ \emph {et~al.}(2003)\citenamefont {Segal},
		\citenamefont {Shapira}, \citenamefont {Regev}, \citenamefont {Pe'er},
		\citenamefont {Botstein}, \citenamefont {Koller},\ and\ \citenamefont
		{Friedman}}]{segal2003module}%
	\BibitemOpen
	\bibfield  {author} {\bibinfo {author} {\bibfnamefont {E.}~\bibnamefont
			{Segal}}, \bibinfo {author} {\bibfnamefont {M.}~\bibnamefont {Shapira}},
		\bibinfo {author} {\bibfnamefont {A.}~\bibnamefont {Regev}}, \bibinfo
		{author} {\bibfnamefont {D.}~\bibnamefont {Pe'er}}, \bibinfo {author}
		{\bibfnamefont {D.}~\bibnamefont {Botstein}}, \bibinfo {author}
		{\bibfnamefont {D.}~\bibnamefont {Koller}}, \ and\ \bibinfo {author}
		{\bibfnamefont {N.}~\bibnamefont {Friedman}},\ }\bibfield  {title} {\enquote
		{\bibinfo {title} {Module networks: identifying regulatory modules and their
				condition-specific regulators from gene expression data},}\ }\href@noop {}
	{\bibfield  {journal} {\bibinfo  {journal} {Nature Genetics}\ }\textbf
		{\bibinfo {volume} {34}},\ \bibinfo {pages} {166--176} (\bibinfo {year}
		{2003})}\BibitemShut {NoStop}%
	\bibitem [{\citenamefont {Newman}(2004)}]{newman2004coauthorship}%
	\BibitemOpen
	\bibfield  {author} {\bibinfo {author} {\bibfnamefont {M.~E.~J.}\
			\bibnamefont {Newman}},\ }\bibfield  {title} {\enquote {\bibinfo {title}
			{Coauthorship networks and patterns of scientific collaboration},}\
	}\href@noop {} {\bibfield  {journal} {\bibinfo  {journal} {Proceedings of the
				National Academy of Sciences}\ }\textbf {\bibinfo {volume} {101}},\ \bibinfo
		{pages} {5200--5205} (\bibinfo {year} {2004})}\BibitemShut {NoStop}%
	\bibitem [{\citenamefont {Menck}\ \emph {et~al.}(2014)\citenamefont {Menck},
		\citenamefont {Heitzig}, \citenamefont {Kurths},\ and\ \citenamefont
		{Joachim~Schellnhuber}}]{menck2014dead}%
	\BibitemOpen
	\bibfield  {author} {\bibinfo {author} {\bibfnamefont {P.~J.}\ \bibnamefont
			{Menck}}, \bibinfo {author} {\bibfnamefont {J.}~\bibnamefont {Heitzig}},
		\bibinfo {author} {\bibfnamefont {J.}~\bibnamefont {Kurths}}, \ and\ \bibinfo
		{author} {\bibfnamefont {H.}~\bibnamefont {Joachim~Schellnhuber}},\
	}\bibfield  {title} {\enquote {\bibinfo {title} {How dead ends undermine
				power grid stability},}\ }\href@noop {} {\bibfield  {journal} {\bibinfo
			{journal} {Nature Communications}\ }\textbf {\bibinfo {volume} {5}},\
		\bibinfo {pages} {3969} (\bibinfo {year} {2014})}\BibitemShut {NoStop}%
	\bibitem [{\citenamefont {Banavar}, \citenamefont {Maritan},\ and\
		\citenamefont {Rinaldo}(1999)}]{banavar1999size}%
	\BibitemOpen
	\bibfield  {author} {\bibinfo {author} {\bibfnamefont {J.~R.}\ \bibnamefont
			{Banavar}}, \bibinfo {author} {\bibfnamefont {A.}~\bibnamefont {Maritan}}, \
		and\ \bibinfo {author} {\bibfnamefont {A.}~\bibnamefont {Rinaldo}},\
	}\bibfield  {title} {\enquote {\bibinfo {title} {Size and form in efficient
				transportation networks},}\ }\href@noop {} {\bibfield  {journal} {\bibinfo
			{journal} {Nature}\ }\textbf {\bibinfo {volume} {399}},\ \bibinfo {pages}
		{130--132} (\bibinfo {year} {1999})}\BibitemShut {NoStop}%
	\bibitem [{\citenamefont {Hecker}\ \emph {et~al.}(2009)\citenamefont {Hecker},
		\citenamefont {Lambeck}, \citenamefont {Toepfer}, \citenamefont
		{Van~Someren},\ and\ \citenamefont {Guthke}}]{hecker2009gene}%
	\BibitemOpen
	\bibfield  {author} {\bibinfo {author} {\bibfnamefont {M.}~\bibnamefont
			{Hecker}}, \bibinfo {author} {\bibfnamefont {S.}~\bibnamefont {Lambeck}},
		\bibinfo {author} {\bibfnamefont {S.}~\bibnamefont {Toepfer}}, \bibinfo
		{author} {\bibfnamefont {E.}~\bibnamefont {Van~Someren}}, \ and\ \bibinfo
		{author} {\bibfnamefont {R.}~\bibnamefont {Guthke}},\ }\bibfield  {title}
	{\enquote {\bibinfo {title} {Gene regulatory network inference: data
				integration in dynamic models—a review},}\ }\href@noop {} {\bibfield
		{journal} {\bibinfo  {journal} {Biosystems}\ }\textbf {\bibinfo {volume}
			{96}},\ \bibinfo {pages} {86--103} (\bibinfo {year} {2009})}\BibitemShut
	{NoStop}%
	\bibitem [{\citenamefont {Boccaletti}\ \emph {et~al.}(2002)\citenamefont
		{Boccaletti}, \citenamefont {Kurths}, \citenamefont {Osipov}, \citenamefont
		{Valladares},\ and\ \citenamefont {Zhou}}]{boccaletti2002synchronization}%
	\BibitemOpen
	\bibfield  {author} {\bibinfo {author} {\bibfnamefont {S.}~\bibnamefont
			{Boccaletti}}, \bibinfo {author} {\bibfnamefont {J.}~\bibnamefont {Kurths}},
		\bibinfo {author} {\bibfnamefont {G.}~\bibnamefont {Osipov}}, \bibinfo
		{author} {\bibfnamefont {D.}~\bibnamefont {Valladares}}, \ and\ \bibinfo
		{author} {\bibfnamefont {C.}~\bibnamefont {Zhou}},\ }\bibfield  {title}
	{\enquote {\bibinfo {title} {The synchronization of chaotic systems},}\
	}\href@noop {} {\bibfield  {journal} {\bibinfo  {journal} {Physics Reports}\
		}\textbf {\bibinfo {volume} {366}},\ \bibinfo {pages} {1--101} (\bibinfo
		{year} {2002})}\BibitemShut {NoStop}%
	\bibitem [{\citenamefont {Pikovsky}, \citenamefont {Rosenblum},\ and\
		\citenamefont {Kurths}(2003)}]{pikovsky2003universal}%
	\BibitemOpen
	\bibfield  {author} {\bibinfo {author} {\bibfnamefont {A.}~\bibnamefont
			{Pikovsky}}, \bibinfo {author} {\bibfnamefont {M.}~\bibnamefont {Rosenblum}},
		\ and\ \bibinfo {author} {\bibfnamefont {J.}~\bibnamefont {Kurths}},\
	}\href@noop {} {\emph {\bibinfo {title} {Synchronization: A Universal Concept
				in Nonlinear Sciences}}}\ (\bibinfo  {publisher} {Cambridge University Press,
		Cambridge},\ \bibinfo {year} {2003})\BibitemShut {NoStop}%
	\bibitem [{\citenamefont {Boccaletti}\ \emph {et~al.}(2018)\citenamefont
		{Boccaletti}, \citenamefont {Pisarchik}, \citenamefont {Del~Genio},\ and\
		\citenamefont {Amann}}]{boccaletti2018synchronization}%
	\BibitemOpen
	\bibfield  {author} {\bibinfo {author} {\bibfnamefont {S.}~\bibnamefont
			{Boccaletti}}, \bibinfo {author} {\bibfnamefont {A.~N.}\ \bibnamefont
			{Pisarchik}}, \bibinfo {author} {\bibfnamefont {C.~I.}\ \bibnamefont
			{Del~Genio}}, \ and\ \bibinfo {author} {\bibfnamefont {A.}~\bibnamefont
			{Amann}},\ }\href@noop {} {\emph {\bibinfo {title} {Synchronization: From
				Coupled Systems to Complex Networks}}}\ (\bibinfo  {publisher} {Cambridge
		University Press},\ \bibinfo {year} {2018})\BibitemShut {NoStop}%
	\bibitem [{\citenamefont {Wu}\ \emph {et~al.}(2024)\citenamefont {Wu},
		\citenamefont {Wu}, \citenamefont {Wang}, \citenamefont {Mao}, \citenamefont
		{Lu}, \citenamefont {L{\"u}}, \citenamefont {Zhang},\ and\ \citenamefont
		{L{\"u}}}]{wu2024synchronization}%
	\BibitemOpen
	\bibfield  {author} {\bibinfo {author} {\bibfnamefont {X.}~\bibnamefont
			{Wu}}, \bibinfo {author} {\bibfnamefont {X.}~\bibnamefont {Wu}}, \bibinfo
		{author} {\bibfnamefont {C.-Y.}\ \bibnamefont {Wang}}, \bibinfo {author}
		{\bibfnamefont {B.}~\bibnamefont {Mao}}, \bibinfo {author} {\bibfnamefont
			{J.-a.}\ \bibnamefont {Lu}}, \bibinfo {author} {\bibfnamefont
			{J.}~\bibnamefont {L{\"u}}}, \bibinfo {author} {\bibfnamefont {Y.-C.}\
			\bibnamefont {Zhang}}, \ and\ \bibinfo {author} {\bibfnamefont
			{L.}~\bibnamefont {L{\"u}}},\ }\bibfield  {title} {\enquote {\bibinfo {title}
			{Synchronization in multiplex networks},}\ }\href@noop {} {\bibfield
		{journal} {\bibinfo  {journal} {Physics Reports}\ }\textbf {\bibinfo {volume}
			{1060}},\ \bibinfo {pages} {1--54} (\bibinfo {year} {2024})}\BibitemShut
	{NoStop}%
	\bibitem [{\citenamefont {Buck}(1988)}]{buck1988synchronous}%
	\BibitemOpen
	\bibfield  {author} {\bibinfo {author} {\bibfnamefont {J.}~\bibnamefont
			{Buck}},\ }\bibfield  {title} {\enquote {\bibinfo {title} {Synchronous
				rhythmic flashing of fireflies},}\ }\href@noop {} {\bibfield  {journal}
		{\bibinfo  {journal} {The Quarterly Review of Biology}\ }\textbf {\bibinfo
			{volume} {63}},\ \bibinfo {pages} {265--289} (\bibinfo {year}
		{1988})}\BibitemShut {NoStop}%
	\bibitem [{\citenamefont {N{\'e}da}\ \emph {et~al.}(2000)\citenamefont
		{N{\'e}da}, \citenamefont {Ravasz}, \citenamefont {Brechet}, \citenamefont
		{Vicsek},\ and\ \citenamefont {Barab{\'a}si}}]{neda2000sound}%
	\BibitemOpen
	\bibfield  {author} {\bibinfo {author} {\bibfnamefont {Z.}~\bibnamefont
			{N{\'e}da}}, \bibinfo {author} {\bibfnamefont {E.}~\bibnamefont {Ravasz}},
		\bibinfo {author} {\bibfnamefont {Y.}~\bibnamefont {Brechet}}, \bibinfo
		{author} {\bibfnamefont {T.}~\bibnamefont {Vicsek}}, \ and\ \bibinfo {author}
		{\bibfnamefont {A.-L.}\ \bibnamefont {Barab{\'a}si}},\ }\bibfield  {title}
	{\enquote {\bibinfo {title} {The sound of many hands clapping},}\ }\href@noop
	{} {\bibfield  {journal} {\bibinfo  {journal} {Nature}\ }\textbf {\bibinfo
			{volume} {403}},\ \bibinfo {pages} {849--850} (\bibinfo {year}
		{2000})}\BibitemShut {NoStop}%
	\bibitem [{\citenamefont {Penn}, \citenamefont {Segal},\ and\ \citenamefont
		{Moses}(2016)}]{penn2016network}%
	\BibitemOpen
	\bibfield  {author} {\bibinfo {author} {\bibfnamefont {Y.}~\bibnamefont
			{Penn}}, \bibinfo {author} {\bibfnamefont {M.}~\bibnamefont {Segal}}, \ and\
		\bibinfo {author} {\bibfnamefont {E.}~\bibnamefont {Moses}},\ }\bibfield
	{title} {\enquote {\bibinfo {title} {Network synchronization in hippocampal
				neurons},}\ }\href@noop {} {\bibfield  {journal} {\bibinfo  {journal}
			{Proceedings of the National Academy of Sciences}\ }\textbf {\bibinfo
			{volume} {113}},\ \bibinfo {pages} {3341--3346} (\bibinfo {year}
		{2016})}\BibitemShut {NoStop}%
	\bibitem [{\citenamefont {De~Monte}\ \emph {et~al.}(2007)\citenamefont
		{De~Monte}, \citenamefont {d'Ovidio}, \citenamefont {Dan{\o}},\ and\
		\citenamefont {S{\o}rensen}}]{de2007dynamical}%
	\BibitemOpen
	\bibfield  {author} {\bibinfo {author} {\bibfnamefont {S.}~\bibnamefont
			{De~Monte}}, \bibinfo {author} {\bibfnamefont {F.}~\bibnamefont {d'Ovidio}},
		\bibinfo {author} {\bibfnamefont {S.}~\bibnamefont {Dan{\o}}}, \ and\
		\bibinfo {author} {\bibfnamefont {P.~G.}\ \bibnamefont {S{\o}rensen}},\
	}\bibfield  {title} {\enquote {\bibinfo {title} {Dynamical quorum sensing:
				Population density encoded in cellular dynamics},}\ }\href@noop {} {\bibfield
		{journal} {\bibinfo  {journal} {Proceedings of the National Academy of
				Sciences}\ }\textbf {\bibinfo {volume} {104}},\ \bibinfo {pages}
		{18377--18381} (\bibinfo {year} {2007})}\BibitemShut {NoStop}%
	\bibitem [{\citenamefont {Motter}\ \emph {et~al.}(2013)\citenamefont {Motter},
		\citenamefont {Myers}, \citenamefont {Anghel},\ and\ \citenamefont
		{Nishikawa}}]{motter2013spontaneous}%
	\BibitemOpen
	\bibfield  {author} {\bibinfo {author} {\bibfnamefont {A.~E.}\ \bibnamefont
			{Motter}}, \bibinfo {author} {\bibfnamefont {S.~A.}\ \bibnamefont {Myers}},
		\bibinfo {author} {\bibfnamefont {M.}~\bibnamefont {Anghel}}, \ and\ \bibinfo
		{author} {\bibfnamefont {T.}~\bibnamefont {Nishikawa}},\ }\bibfield  {title}
	{\enquote {\bibinfo {title} {Spontaneous synchrony in power-grid networks},}\
	}\href@noop {} {\bibfield  {journal} {\bibinfo  {journal} {Nature Physics}\
		}\textbf {\bibinfo {volume} {9}},\ \bibinfo {pages} {191--197} (\bibinfo
		{year} {2013})}\BibitemShut {NoStop}%
	\bibitem [{\citenamefont {Kuramoto}(1984)}]{kuramoto1984chemical}%
	\BibitemOpen
	\bibfield  {author} {\bibinfo {author} {\bibfnamefont {Y.}~\bibnamefont
			{Kuramoto}},\ }\href@noop {} {\emph {\bibinfo {title} {Chemical Oscillations,
				Waves and Turbulence}}}\ (\bibinfo  {publisher} {Springer, New York},\
	\bibinfo {year} {1984})\BibitemShut {NoStop}%
	\bibitem [{\citenamefont {G{\'o}mez-Gardenes}, \citenamefont {Moreno},\ and\
		\citenamefont {Arenas}(2007)}]{gomez2007paths}%
	\BibitemOpen
	\bibfield  {author} {\bibinfo {author} {\bibfnamefont {J.}~\bibnamefont
			{G{\'o}mez-Gardenes}}, \bibinfo {author} {\bibfnamefont {Y.}~\bibnamefont
			{Moreno}}, \ and\ \bibinfo {author} {\bibfnamefont {A.}~\bibnamefont
			{Arenas}},\ }\bibfield  {title} {\enquote {\bibinfo {title} {Paths to
				synchronization on complex networks},}\ }\href@noop {} {\bibfield  {journal}
		{\bibinfo  {journal} {Physical Review Letters}\ }\textbf {\bibinfo {volume}
			{98}},\ \bibinfo {pages} {034101} (\bibinfo {year} {2007})}\BibitemShut
	{NoStop}%
	\bibitem [{\citenamefont {G{\'o}mez-Gardenes}\ \emph
		{et~al.}(2011)\citenamefont {G{\'o}mez-Gardenes}, \citenamefont {G{\'o}mez},
		\citenamefont {Arenas},\ and\ \citenamefont {Moreno}}]{gomez2011explosive}%
	\BibitemOpen
	\bibfield  {author} {\bibinfo {author} {\bibfnamefont {J.}~\bibnamefont
			{G{\'o}mez-Gardenes}}, \bibinfo {author} {\bibfnamefont {S.}~\bibnamefont
			{G{\'o}mez}}, \bibinfo {author} {\bibfnamefont {A.}~\bibnamefont {Arenas}}, \
		and\ \bibinfo {author} {\bibfnamefont {Y.}~\bibnamefont {Moreno}},\
	}\bibfield  {title} {\enquote {\bibinfo {title} {Explosive synchronization
				transitions in scale-free networks},}\ }\href@noop {} {\bibfield  {journal}
		{\bibinfo  {journal} {Physical Review Letters}\ }\textbf {\bibinfo {volume}
			{106}},\ \bibinfo {pages} {128701} (\bibinfo {year} {2011})}\BibitemShut
	{NoStop}%
	\bibitem [{\citenamefont {Zhang}\ \emph {et~al.}(2015)\citenamefont {Zhang},
		\citenamefont {Boccaletti}, \citenamefont {Guan},\ and\ \citenamefont
		{Liu}}]{zhang2015explosive}%
	\BibitemOpen
	\bibfield  {author} {\bibinfo {author} {\bibfnamefont {X.}~\bibnamefont
			{Zhang}}, \bibinfo {author} {\bibfnamefont {S.}~\bibnamefont {Boccaletti}},
		\bibinfo {author} {\bibfnamefont {S.}~\bibnamefont {Guan}}, \ and\ \bibinfo
		{author} {\bibfnamefont {Z.}~\bibnamefont {Liu}},\ }\bibfield  {title}
	{\enquote {\bibinfo {title} {Explosive synchronization in adaptive and
				multilayer networks},}\ }\href@noop {} {\bibfield  {journal} {\bibinfo
			{journal} {Physical Review Letters}\ }\textbf {\bibinfo {volume} {114}},\
		\bibinfo {pages} {038701} (\bibinfo {year} {2015})}\BibitemShut {NoStop}%
	\bibitem [{\citenamefont {Kumar}, \citenamefont {Jalan},\ and\ \citenamefont
		{Kachhvah}(2020)}]{kumar2020interlayer}%
	\BibitemOpen
	\bibfield  {author} {\bibinfo {author} {\bibfnamefont {A.}~\bibnamefont
			{Kumar}}, \bibinfo {author} {\bibfnamefont {S.}~\bibnamefont {Jalan}}, \ and\
		\bibinfo {author} {\bibfnamefont {A.~D.}\ \bibnamefont {Kachhvah}},\
	}\bibfield  {title} {\enquote {\bibinfo {title} {Interlayer
				adaptation-induced explosive synchronization in multiplex networks},}\
	}\href@noop {} {\bibfield  {journal} {\bibinfo  {journal} {Physical Review
				Research}\ }\textbf {\bibinfo {volume} {2}},\ \bibinfo {pages} {023259}
		(\bibinfo {year} {2020})}\BibitemShut {NoStop}%
	\bibitem [{\citenamefont {Wu}\ \emph {et~al.}(2022)\citenamefont {Wu},
		\citenamefont {Huo}, \citenamefont {Alfaro-Bittner}, \citenamefont
		{Boccaletti},\ and\ \citenamefont {Liu}}]{wu2022double}%
	\BibitemOpen
	\bibfield  {author} {\bibinfo {author} {\bibfnamefont {T.}~\bibnamefont
			{Wu}}, \bibinfo {author} {\bibfnamefont {S.}~\bibnamefont {Huo}}, \bibinfo
		{author} {\bibfnamefont {K.}~\bibnamefont {Alfaro-Bittner}}, \bibinfo
		{author} {\bibfnamefont {S.}~\bibnamefont {Boccaletti}}, \ and\ \bibinfo
		{author} {\bibfnamefont {Z.}~\bibnamefont {Liu}},\ }\bibfield  {title}
	{\enquote {\bibinfo {title} {Double explosive transition in the
				synchronization of multilayer networks},}\ }\href@noop {} {\bibfield
		{journal} {\bibinfo  {journal} {Physical Review Research}\ }\textbf {\bibinfo
			{volume} {4}},\ \bibinfo {pages} {033009} (\bibinfo {year}
		{2022})}\BibitemShut {NoStop}%
	\bibitem [{\citenamefont {Pal}\ \emph {et~al.}(2025)\citenamefont {Pal},
		\citenamefont {Frolov}, \citenamefont {Rakshit}, \citenamefont {Hramov},\
		and\ \citenamefont {Ghosh}}]{pal2025explosive}%
	\BibitemOpen
	\bibfield  {author} {\bibinfo {author} {\bibfnamefont {P.~K.}\ \bibnamefont
			{Pal}}, \bibinfo {author} {\bibfnamefont {N.}~\bibnamefont {Frolov}},
		\bibinfo {author} {\bibfnamefont {S.}~\bibnamefont {Rakshit}}, \bibinfo
		{author} {\bibfnamefont {A.~E.}\ \bibnamefont {Hramov}}, \ and\ \bibinfo
		{author} {\bibfnamefont {D.}~\bibnamefont {Ghosh}},\ }\bibfield  {title}
	{\enquote {\bibinfo {title} {Explosive synchronization in generalized
				multiplex network with competitive and cooperative interlayer
				interactions},}\ }\href@noop {} {\bibfield  {journal} {\bibinfo  {journal}
			{Chaos: An Interdisciplinary Journal of Nonlinear Science}\ }\textbf
		{\bibinfo {volume} {35}},\ \bibinfo {pages} {071102} (\bibinfo {year}
		{2025})}\BibitemShut {NoStop}%
	\bibitem [{\citenamefont {Sakaguchi}\ and\ \citenamefont
		{Kuramoto}(1986)}]{sakaguchi1986soluble}%
	\BibitemOpen
	\bibfield  {author} {\bibinfo {author} {\bibfnamefont {H.}~\bibnamefont
			{Sakaguchi}}\ and\ \bibinfo {author} {\bibfnamefont {Y.}~\bibnamefont
			{Kuramoto}},\ }\bibfield  {title} {\enquote {\bibinfo {title} {A soluble
				active rotater model showing phase transitions via mutual entertainment},}\
	}\href@noop {} {\bibfield  {journal} {\bibinfo  {journal} {Progress of
				Theoretical Physics}\ }\textbf {\bibinfo {volume} {76}},\ \bibinfo {pages}
		{576--581} (\bibinfo {year} {1986})}\BibitemShut {NoStop}%
	\bibitem [{\citenamefont {Wiesenfeld}, \citenamefont {Colet},\ and\
		\citenamefont {Strogatz}(1998)}]{wiesenfeld1998frequency}%
	\BibitemOpen
	\bibfield  {author} {\bibinfo {author} {\bibfnamefont {K.}~\bibnamefont
			{Wiesenfeld}}, \bibinfo {author} {\bibfnamefont {P.}~\bibnamefont {Colet}}, \
		and\ \bibinfo {author} {\bibfnamefont {S.~H.}\ \bibnamefont {Strogatz}},\
	}\bibfield  {title} {\enquote {\bibinfo {title} {Frequency locking in
				\text{Josephson} arrays: Connection with the \text{Kuramoto} model},}\
	}\href@noop {} {\bibfield  {journal} {\bibinfo  {journal} {Physical Review
				E}\ }\textbf {\bibinfo {volume} {57}},\ \bibinfo {pages} {1563} (\bibinfo
		{year} {1998})}\BibitemShut {NoStop}%
	\bibitem [{\citenamefont {Dorfler}\ and\ \citenamefont
		{Bullo}(2012)}]{dorfler2012synchronization}%
	\BibitemOpen
	\bibfield  {author} {\bibinfo {author} {\bibfnamefont {F.}~\bibnamefont
			{Dorfler}}\ and\ \bibinfo {author} {\bibfnamefont {F.}~\bibnamefont
			{Bullo}},\ }\bibfield  {title} {\enquote {\bibinfo {title} {Synchronization
				and transient stability in power networks and nonuniform \text{Kuramoto}
				oscillators},}\ }\href@noop {} {\bibfield  {journal} {\bibinfo  {journal}
			{SIAM Journal on Control and Optimization}\ }\textbf {\bibinfo {volume}
			{50}},\ \bibinfo {pages} {1616--1642} (\bibinfo {year} {2012})}\BibitemShut
	{NoStop}%
	\bibitem [{\citenamefont {Abrams}\ and\ \citenamefont
		{Strogatz}(2004)}]{abrams2004chimera}%
	\BibitemOpen
	\bibfield  {author} {\bibinfo {author} {\bibfnamefont {D.~M.}\ \bibnamefont
			{Abrams}}\ and\ \bibinfo {author} {\bibfnamefont {S.~H.}\ \bibnamefont
			{Strogatz}},\ }\bibfield  {title} {\enquote {\bibinfo {title} {Chimera states
				for coupled oscillators},}\ }\href@noop {} {\bibfield  {journal} {\bibinfo
			{journal} {Physical Review Letters}\ }\textbf {\bibinfo {volume} {93}},\
		\bibinfo {pages} {174102} (\bibinfo {year} {2004})}\BibitemShut {NoStop}%
	\bibitem [{\citenamefont {Kundu}\ \emph
		{et~al.}(2017{\natexlab{a}})\citenamefont {Kundu}, \citenamefont {Khanra},
		\citenamefont {Hens},\ and\ \citenamefont {Pal}}]{kundu2017transition}%
	\BibitemOpen
	\bibfield  {author} {\bibinfo {author} {\bibfnamefont {P.}~\bibnamefont
			{Kundu}}, \bibinfo {author} {\bibfnamefont {P.}~\bibnamefont {Khanra}},
		\bibinfo {author} {\bibfnamefont {C.}~\bibnamefont {Hens}}, \ and\ \bibinfo
		{author} {\bibfnamefont {P.}~\bibnamefont {Pal}},\ }\bibfield  {title}
	{\enquote {\bibinfo {title} {Transition to synchrony in degree-frequency
				correlated \text{Sakaguchi-Kuramoto} model},}\ }\href@noop {} {\bibfield
		{journal} {\bibinfo  {journal} {Physical Review E}\ }\textbf {\bibinfo
			{volume} {96}},\ \bibinfo {pages} {052216} (\bibinfo {year}
		{2017}{\natexlab{a}})}\BibitemShut {NoStop}%
	\bibitem [{\citenamefont {Khanra}\ \emph {et~al.}(2018)\citenamefont {Khanra},
		\citenamefont {Kundu}, \citenamefont {Hens},\ and\ \citenamefont
		{Pal}}]{khanra2018explosive}%
	\BibitemOpen
	\bibfield  {author} {\bibinfo {author} {\bibfnamefont {P.}~\bibnamefont
			{Khanra}}, \bibinfo {author} {\bibfnamefont {P.}~\bibnamefont {Kundu}},
		\bibinfo {author} {\bibfnamefont {C.}~\bibnamefont {Hens}}, \ and\ \bibinfo
		{author} {\bibfnamefont {P.}~\bibnamefont {Pal}},\ }\bibfield  {title}
	{\enquote {\bibinfo {title} {Explosive synchronization in phase-frustrated
				multiplex networks},}\ }\href@noop {} {\bibfield  {journal} {\bibinfo
			{journal} {Physical Review E}\ }\textbf {\bibinfo {volume} {98}},\ \bibinfo
		{pages} {052315} (\bibinfo {year} {2018})}\BibitemShut {NoStop}%
	\bibitem [{\citenamefont {Kumar}\ and\ \citenamefont
		{Jalan}(2021)}]{kumar2021explosive}%
	\BibitemOpen
	\bibfield  {author} {\bibinfo {author} {\bibfnamefont {A.}~\bibnamefont
			{Kumar}}\ and\ \bibinfo {author} {\bibfnamefont {S.}~\bibnamefont {Jalan}},\
	}\bibfield  {title} {\enquote {\bibinfo {title} {Explosive synchronization in
				interlayer phase-shifted \text{Kuramoto} oscillators on multiplex
				networks},}\ }\href@noop {} {\bibfield  {journal} {\bibinfo  {journal}
			{Chaos: An Interdisciplinary Journal of Nonlinear Science}\ }\textbf
		{\bibinfo {volume} {31}},\ \bibinfo {pages} {041103} (\bibinfo {year}
		{2021})}\BibitemShut {NoStop}%
	\bibitem [{\citenamefont {Seif}\ and\ \citenamefont
		{Zarei}(2025)}]{seif2025double}%
	\BibitemOpen
	\bibfield  {author} {\bibinfo {author} {\bibfnamefont {A.}~\bibnamefont
			{Seif}}\ and\ \bibinfo {author} {\bibfnamefont {M.}~\bibnamefont {Zarei}},\
	}\bibfield  {title} {\enquote {\bibinfo {title} {Double hysteresis loop in
				synchronization transitions of multiplex networks: The role of frequency
				arrangements and frustration},}\ }\href@noop {} {\bibfield  {journal}
		{\bibinfo  {journal} {Chaos, Solitons \& Fractals}\ }\textbf {\bibinfo
			{volume} {196}},\ \bibinfo {pages} {116412} (\bibinfo {year}
		{2025})}\BibitemShut {NoStop}%
	\bibitem [{\citenamefont {Chen}(2022)}]{chen2022searching}%
	\BibitemOpen
	\bibfield  {author} {\bibinfo {author} {\bibfnamefont {G.}~\bibnamefont
			{Chen}},\ }\bibfield  {title} {\enquote {\bibinfo {title} {Searching for best
				network topologies with optimal synchronizability: A brief review},}\
	}\href@noop {} {\bibfield  {journal} {\bibinfo  {journal} {IEEE-CAA Journal
				of Automatica Sinica}\ }\textbf {\bibinfo {volume} {9}},\ \bibinfo {pages}
		{573--577} (\bibinfo {year} {2022})}\BibitemShut {NoStop}%
	\bibitem [{\citenamefont {Brede}(2008)}]{brede2008synchrony}%
	\BibitemOpen
	\bibfield  {author} {\bibinfo {author} {\bibfnamefont {M.}~\bibnamefont
			{Brede}},\ }\bibfield  {title} {\enquote {\bibinfo {title}
			{Synchrony-optimized networks of non-identical \text{Kuramoto}
				oscillators},}\ }\href@noop {} {\bibfield  {journal} {\bibinfo  {journal}
			{Physics Letters A}\ }\textbf {\bibinfo {volume} {372}},\ \bibinfo {pages}
		{2618--2622} (\bibinfo {year} {2008})}\BibitemShut {NoStop}%
	\bibitem [{\citenamefont {Skardal}, \citenamefont {Taylor},\ and\ \citenamefont
		{Sun}(2014)}]{skardal2014optimal}%
	\BibitemOpen
	\bibfield  {author} {\bibinfo {author} {\bibfnamefont {P.~S.}\ \bibnamefont
			{Skardal}}, \bibinfo {author} {\bibfnamefont {D.}~\bibnamefont {Taylor}}, \
		and\ \bibinfo {author} {\bibfnamefont {J.}~\bibnamefont {Sun}},\ }\bibfield
	{title} {\enquote {\bibinfo {title} {Optimal synchronization of complex
				networks},}\ }\href@noop {} {\bibfield  {journal} {\bibinfo  {journal}
			{Physical Review Letters}\ }\textbf {\bibinfo {volume} {113}},\ \bibinfo
		{pages} {144101} (\bibinfo {year} {2014})}\BibitemShut {NoStop}%
	\bibitem [{\citenamefont {Ott}\ and\ \citenamefont
		{Antonsen}(2008)}]{ott2008low}%
	\BibitemOpen
	\bibfield  {author} {\bibinfo {author} {\bibfnamefont {E.}~\bibnamefont
			{Ott}}\ and\ \bibinfo {author} {\bibfnamefont {T.~M.}\ \bibnamefont
			{Antonsen}},\ }\bibfield  {title} {\enquote {\bibinfo {title} {Low
				dimensional behavior of large systems of globally coupled oscillators},}\
	}\href@noop {} {\bibfield  {journal} {\bibinfo  {journal} {Chaos: An
				Interdisciplinary Journal of Nonlinear Science}\ }\textbf {\bibinfo {volume}
			{18}},\ \bibinfo {pages} {037113} (\bibinfo {year} {2008})}\BibitemShut
	{NoStop}%
	\bibitem [{\citenamefont {Gottwald}(2015)}]{gottwald2015model}%
	\BibitemOpen
	\bibfield  {author} {\bibinfo {author} {\bibfnamefont {G.~A.}\ \bibnamefont
			{Gottwald}},\ }\bibfield  {title} {\enquote {\bibinfo {title} {Model
				reduction for networks of coupled oscillators},}\ }\href@noop {} {\bibfield
		{journal} {\bibinfo  {journal} {Chaos: An Interdisciplinary Journal of
				Nonlinear Science}\ }\textbf {\bibinfo {volume} {25}},\ \bibinfo {pages}
		{053111} (\bibinfo {year} {2015})}\BibitemShut {NoStop}%
	\bibitem [{\citenamefont {Pinto}\ and\ \citenamefont
		{Saa}(2015)}]{pinto2015optimal}%
	\BibitemOpen
	\bibfield  {author} {\bibinfo {author} {\bibfnamefont {R.~S.}\ \bibnamefont
			{Pinto}}\ and\ \bibinfo {author} {\bibfnamefont {A.}~\bibnamefont {Saa}},\
	}\bibfield  {title} {\enquote {\bibinfo {title} {Optimal synchronization of
				\text{Kuramoto} oscillators: A dimensional reduction approach},}\ }\href@noop
	{} {\bibfield  {journal} {\bibinfo  {journal} {Physical Review E}\ }\textbf
		{\bibinfo {volume} {92}},\ \bibinfo {pages} {062801} (\bibinfo {year}
		{2015})}\BibitemShut {NoStop}%
	\bibitem [{\citenamefont {Skardal}, \citenamefont {Taylor},\ and\ \citenamefont
		{Sun}(2016)}]{skardal2016optimal}%
	\BibitemOpen
	\bibfield  {author} {\bibinfo {author} {\bibfnamefont {P.~S.}\ \bibnamefont
			{Skardal}}, \bibinfo {author} {\bibfnamefont {D.}~\bibnamefont {Taylor}}, \
		and\ \bibinfo {author} {\bibfnamefont {J.}~\bibnamefont {Sun}},\ }\bibfield
	{title} {\enquote {\bibinfo {title} {Optimal synchronization of directed
				complex networks},}\ }\href@noop {} {\bibfield  {journal} {\bibinfo
			{journal} {Chaos: An Interdisciplinary Journal of Nonlinear Science}\
		}\textbf {\bibinfo {volume} {26}},\ \bibinfo {pages} {094807} (\bibinfo
		{year} {2016})}\BibitemShut {NoStop}%
	\bibitem [{\citenamefont {Kundu}\ \emph
		{et~al.}(2017{\natexlab{b}})\citenamefont {Kundu}, \citenamefont {Hens},
		\citenamefont {Barzel},\ and\ \citenamefont {Pal}}]{kundu2017perfect}%
	\BibitemOpen
	\bibfield  {author} {\bibinfo {author} {\bibfnamefont {P.}~\bibnamefont
			{Kundu}}, \bibinfo {author} {\bibfnamefont {C.}~\bibnamefont {Hens}},
		\bibinfo {author} {\bibfnamefont {B.}~\bibnamefont {Barzel}}, \ and\ \bibinfo
		{author} {\bibfnamefont {P.}~\bibnamefont {Pal}},\ }\bibfield  {title}
	{\enquote {\bibinfo {title} {Perfect synchronization in networks of
				phase-frustrated oscillators},}\ }\href@noop {} {\bibfield  {journal}
		{\bibinfo  {journal} {Europhysics Letters}\ }\textbf {\bibinfo {volume}
			{120}},\ \bibinfo {pages} {40002} (\bibinfo {year}
		{2017}{\natexlab{b}})}\BibitemShut {NoStop}%
	\bibitem [{\citenamefont {Kundu}\ \emph {et~al.}(2020)\citenamefont {Kundu},
		\citenamefont {Khanra}, \citenamefont {Hens},\ and\ \citenamefont
		{Pal}}]{kundu2020optimizing}%
	\BibitemOpen
	\bibfield  {author} {\bibinfo {author} {\bibfnamefont {P.}~\bibnamefont
			{Kundu}}, \bibinfo {author} {\bibfnamefont {P.}~\bibnamefont {Khanra}},
		\bibinfo {author} {\bibfnamefont {C.}~\bibnamefont {Hens}}, \ and\ \bibinfo
		{author} {\bibfnamefont {P.}~\bibnamefont {Pal}},\ }\bibfield  {title}
	{\enquote {\bibinfo {title} {Optimizing synchronization in multiplex networks
				of phase oscillators},}\ }\href@noop {} {\bibfield  {journal} {\bibinfo
			{journal} {Europhysics Letters}\ }\textbf {\bibinfo {volume} {129}},\
		\bibinfo {pages} {30004} (\bibinfo {year} {2020})}\BibitemShut {NoStop}%
	\bibitem [{\citenamefont {Das}, \citenamefont {Kundu},\ and\ \citenamefont
		{Pal}(2025)}]{das2025perfect}%
	\BibitemOpen
	\bibfield  {author} {\bibinfo {author} {\bibfnamefont {A.~B.}\ \bibnamefont
			{Das}}, \bibinfo {author} {\bibfnamefont {P.}~\bibnamefont {Kundu}}, \ and\
		\bibinfo {author} {\bibfnamefont {P.}~\bibnamefont {Pal}},\ }\bibfield
	{title} {\enquote {\bibinfo {title} {Perfect synchronization in
				\text{Sakaguchi--Kuramoto} model on directed complex networks},}\ }\href@noop
	{} {\bibfield  {journal} {\bibinfo  {journal} {Chaos, Solitons \& Fractals}\
		}\textbf {\bibinfo {volume} {198}},\ \bibinfo {pages} {116486} (\bibinfo
		{year} {2025})}\BibitemShut {NoStop}%
	\bibitem [{\citenamefont {Arenas}\ \emph {et~al.}(2008)\citenamefont {Arenas},
		\citenamefont {D{\'\i}az-Guilera}, \citenamefont {Kurths}, \citenamefont
		{Moreno},\ and\ \citenamefont {Zhou}}]{arenas2008synchronization}%
	\BibitemOpen
	\bibfield  {author} {\bibinfo {author} {\bibfnamefont {A.}~\bibnamefont
			{Arenas}}, \bibinfo {author} {\bibfnamefont {A.}~\bibnamefont
			{D{\'\i}az-Guilera}}, \bibinfo {author} {\bibfnamefont {J.}~\bibnamefont
			{Kurths}}, \bibinfo {author} {\bibfnamefont {Y.}~\bibnamefont {Moreno}}, \
		and\ \bibinfo {author} {\bibfnamefont {C.}~\bibnamefont {Zhou}},\ }\bibfield
	{title} {\enquote {\bibinfo {title} {Synchronization in complex networks},}\
	}\href@noop {} {\bibfield  {journal} {\bibinfo  {journal} {Physics Reports}\
		}\textbf {\bibinfo {volume} {469}},\ \bibinfo {pages} {93--153} (\bibinfo
		{year} {2008})}\BibitemShut {NoStop}%
	\bibitem [{\citenamefont {Ben-Israel}\ and\ \citenamefont
		{Grenville}(1974)}]{ben1974generalized}%
	\BibitemOpen
	\bibfield  {author} {\bibinfo {author} {\bibfnamefont {A.}~\bibnamefont
			{Ben-Israel}}\ and\ \bibinfo {author} {\bibfnamefont {T.~N.~E.}\ \bibnamefont
			{Grenville}},\ }\href@noop {} {\emph {\bibinfo {title} {Generalized
				Inverses}}}\ (\bibinfo  {publisher} {Springer, New York},\ \bibinfo {year}
	{1974})\BibitemShut {NoStop}%
	\bibitem [{\citenamefont {Golub}\ and\ \citenamefont
		{Van~Loan}(1996)}]{golub1996matrix}%
	\BibitemOpen
	\bibfield  {author} {\bibinfo {author} {\bibfnamefont {G.~H.}\ \bibnamefont
			{Golub}}\ and\ \bibinfo {author} {\bibfnamefont {C.~F.}\ \bibnamefont
			{Van~Loan}},\ }\href@noop {} {\emph {\bibinfo {title} {Matrix
				Computations}}}\ (\bibinfo  {publisher} {The John Hopkins University Press},\
	\bibinfo {year} {1996})\BibitemShut {NoStop}%
	\bibitem [{\citenamefont {Bollob{\'a}s}\ \emph {et~al.}(2003)\citenamefont
		{Bollob{\'a}s}, \citenamefont {Borgs}, \citenamefont {Chayes},\ and\
		\citenamefont {Riordan}}]{bollobas2003directed}%
	\BibitemOpen
	\bibfield  {author} {\bibinfo {author} {\bibfnamefont {B.}~\bibnamefont
			{Bollob{\'a}s}}, \bibinfo {author} {\bibfnamefont {C.}~\bibnamefont {Borgs}},
		\bibinfo {author} {\bibfnamefont {J.~T.}\ \bibnamefont {Chayes}}, \ and\
		\bibinfo {author} {\bibfnamefont {O.}~\bibnamefont {Riordan}},\ }\bibfield
	{title} {\enquote {\bibinfo {title} {Directed scale-free graphs},}\ }in\
	\href@noop {} {\emph {\bibinfo {booktitle} {Proceedings of the Fourteenth
				Annual ACM-SIAM Symposium on Discrete Algorithms}}},\ Vol.~\bibinfo {volume}
	{3}\ (\bibinfo {organization} {ACM},\ \bibinfo {year} {2003})\ pp.\ \bibinfo
	{pages} {132--139}\BibitemShut {NoStop}%
	\bibitem [{\citenamefont {Erd{\H{o}}s}\ and\ \citenamefont
		{R{\'e}nyi}(1959)}]{erdos1959random}%
	\BibitemOpen
	\bibfield  {author} {\bibinfo {author} {\bibfnamefont {P.}~\bibnamefont
			{Erd{\H{o}}s}}\ and\ \bibinfo {author} {\bibfnamefont {A.}~\bibnamefont
			{R{\'e}nyi}},\ }\bibfield  {title} {\enquote {\bibinfo {title} {On random
				graphs},}\ }\href@noop {} {\bibfield  {journal} {\bibinfo  {journal}
			{Publicationes Mathematicae Debrecen}\ }\textbf {\bibinfo {volume} {6}},\
		\bibinfo {pages} {290--297} (\bibinfo {year} {1959})}\BibitemShut {NoStop}%
	\bibitem [{\citenamefont {Gilbert}(1959)}]{gilbert1959random}%
	\BibitemOpen
	\bibfield  {author} {\bibinfo {author} {\bibfnamefont {E.~N.}\ \bibnamefont
			{Gilbert}},\ }\bibfield  {title} {\enquote {\bibinfo {title} {Random
				graphs},}\ }\href@noop {} {\bibfield  {journal} {\bibinfo  {journal} {The
				Annals of Mathematical Statistics}\ }\textbf {\bibinfo {volume} {30}},\
		\bibinfo {pages} {1141--1144} (\bibinfo {year} {1959})}\BibitemShut {NoStop}%
	\bibitem [{\citenamefont {Bj{\"o}rner}\ and\ \citenamefont
		{Welker}(1999)}]{bjorner1999complexes}%
	\BibitemOpen
	\bibfield  {author} {\bibinfo {author} {\bibfnamefont {A.}~\bibnamefont
			{Bj{\"o}rner}}\ and\ \bibinfo {author} {\bibfnamefont {V.}~\bibnamefont
			{Welker}},\ }\bibfield  {title} {\enquote {\bibinfo {title} {Complexes of
				directed graphs},}\ }\href@noop {} {\bibfield  {journal} {\bibinfo  {journal}
			{SIAM Journal on Discrete Mathematics}\ }\textbf {\bibinfo {volume} {12}},\
		\bibinfo {pages} {413--424} (\bibinfo {year} {1999})}\BibitemShut {NoStop}%
	\bibitem [{\citenamefont {Gallo}\ \emph {et~al.}(1993)\citenamefont {Gallo},
		\citenamefont {Longo}, \citenamefont {Pallottino},\ and\ \citenamefont
		{Nguyen}}]{gallo1993directed}%
	\BibitemOpen
	\bibfield  {author} {\bibinfo {author} {\bibfnamefont {G.}~\bibnamefont
			{Gallo}}, \bibinfo {author} {\bibfnamefont {G.}~\bibnamefont {Longo}},
		\bibinfo {author} {\bibfnamefont {S.}~\bibnamefont {Pallottino}}, \ and\
		\bibinfo {author} {\bibfnamefont {S.}~\bibnamefont {Nguyen}},\ }\bibfield
	{title} {\enquote {\bibinfo {title} {Directed hypergraphs and
				applications},}\ }\href@noop {} {\bibfield  {journal} {\bibinfo  {journal}
			{Discrete Applied Mathematics}\ }\textbf {\bibinfo {volume} {42}},\ \bibinfo
		{pages} {177--201} (\bibinfo {year} {1993})}\BibitemShut {NoStop}%
\end{thebibliography}
\end{document}